\documentclass[a4paper,11pt]{article}
\usepackage[compact]{titlesec}
\usepackage[left=2.5cm,top=2.5cm,right=2.5cm,bottom=2.5cm]{geometry}
\usepackage[numbers,elide,sort&compress]{natbib}
\usepackage{amssymb,amsmath,bm}
\usepackage{bbold}
\usepackage{appendix}
\usepackage{cleveref}
\usepackage{setspace}

\usepackage{amsmath}
\usepackage{dcolumn}
\usepackage[utf8]{inputenc}
\usepackage{graphics,graphicx}
\usepackage{float}

\pdfoutput=1

\usepackage{color}

\newcommand{\p}{^\prime}
\newcommand{\pp}{^{\prime\prime}}

\usepackage{authblk}

\begin{document}

\title{Accurate Prediction of the Ammonia Probes of a Variable Proton-to-Electron Mass Ratio\\[2mm]}

\author[1,2]{A. Owens}
\author[2]{S. N. Yurchenko}
\author[1]{W. Thiel}
\author[3,4]{V. \v{S}pirko\thanks{The corresponding author: \texttt{spirko@marge.uochb.cas.cz}}}
\affil[1]{\small\emph{ Max-Planck-Institut f\"{u}r Kohlenforschung, Kaiser-Wilhelm-Platz 1, 45470 M\"{u}lheim an der Ruhr, Germany}}
\affil[2]{\emph{Department of Physics and Astronomy, University College London, Gower Street, WC1E 6BT London, United Kingdom}}
\affil[3]{\emph{Academy of Sciences of the Czech Republic, Institute of Organic Chemistry and Biochemistry, Flemingovo n\'am.~2, 166 10 Prague 6, Czech Republic}}
\affil[4]{\emph{Department of Chemical Physics and Optics, Faculty of Mathematics and Physics, Charles University in Prague, Ke Karlovu 3, CZ-12116 Prague 2, Czech Republic}}

\renewcommand\Authands{ and }

\date{}

\onehalfspacing
\maketitle

\label{firstpage}

\doublespacing

\begin{abstract}
\normalsize A comprehensive study of the mass sensitivity of the vibration-rotation-inversion transitions of $^{14}$NH$_3$, $^{15}$NH$_3$, $^{14}$ND$_3$, and $^{15}$ND$_3$ is carried out variationally using the TROVE approach. Variational calculations are robust and accurate, offering a new way to compute sensitivity coefficients. Particular attention is paid to the $\Delta k\!=\!\pm 3$ transitions between the accidentally coinciding rotation-inversion energy levels of the $\nu_2=0^+,0^-,1^+$ and $1^-$ states, and the inversion transitions in the $\nu_4=1$ state affected by the ``giant'' $l$-type doubling effect. These transitions exhibit highly anomalous sensitivities, thus appearing as promising probes of a possible cosmological variation of the proton-to-electron mass ratio $\mu$. Moreover, a simultaneous comparison of the calculated sensitivities reveals a sizeable isotopic dependence which could aid an exclusive ammonia detection.
\end{abstract}

%\begin{keywords}
%molecular data - infrared: ISM - submillimetre: ISM - cosmological parameters
%\end{keywords}

\newpage

\section{Introduction}

Molecular spectroscopy is a well established discipline and the increasing precision of measurements has provided the capacity to test fundamental physics. Recently, a set of several ``forbidden'' $\Delta k\!=\!\pm 3$ transitions between the rotation-inversion energy levels of $^{14}$NH$_3$ in the $\nu_2$ vibrational state were proposed as a promising tool to probe a possible space-time variation of the proton-to-electron mass ratio $\mu=m_p/m_e$~\citep{Jansen:2014,Spirko:2014}. The anomalous mass dependency of these transitions arises from accidental near-degeneracies of the involved energy levels. The sensitivity coefficient $T_{u,l}$, defined as
\begin{equation}
T_{u,l}=\frac{\mu}{E_u-E_l}\Bigl(\frac{{\rm d}E_u}{{\rm d}\mu}-\frac{{\rm d}E_l}{{\rm d}\mu}\Bigr),
\label{eq.T}
\end{equation}
where $E_u$ and $E_l$ refer to the energy of the upper and lower state respectively, allows one to quantify the effect that a possible variation of $\mu$ would have for a given transition. The larger the magnitude of $T_{u,l}$, the more favourable a transition is to test for a drifting $\mu$.

The so-called ammonia method~\citep{Flambaum:2007}, which was adapted from \citet{Veldhoven:2004}, relies on the inversion transitions in the vibrational ground state of $^{14}$NH$_3$. Constraints on a temporal variation of $\mu$ have been determined using this method from measurements of the object B0218$+$357 at redshift $z\sim 0.685$~\citep{Flambaum:2007,Murphy:2008,Kanekar:2011}, and of the system PKS1830$-$211 at $z\sim 0.886$~\citep{Henkel:2009}. A major source of systematic error when using the ammonia method is the comparison with rotational lines from other molecular species, particularly molecules that are non-nitrogen-bearing (see \citet{Murphy:2008}, \citet{Henkel:2009}, and \citet{Kanekar:2011} for a more complete discussion). The most stringent limit using ammonia~\citep{Kanekar:2011} has since been improved upon with methanol absorption spectra observed in the lensing galaxy PKS1830$-$211~\citep{Bagdonaite:2013b}. Three different radio telescopes were used to measure ten absorption lines with sensitivity coefficients ranging from $T=-1.0$ to $-32.8$.

Here we present a comprehensive study of the mass sensitivity of the vibration-rotation-inversion transitions of $^{14}$NH$_3$, $^{15}$NH$_3$, $^{14}$ND$_3$, and $^{15}$ND$_3$. A joint comparison of all relevant isotopic transitions could open the door to an all-ammonia detection, and potentially eliminate certain systematic errors that arise from using alternative reference molecules. We also note that the transitions of the $^{15}$N isotopes are optically thin and free of the nuclear quadrupole structures, providing a simpler radiative and line-shape analysis. A rigorous evaluation of the sensitivity coefficients will hopefully offer new scope for the ammonia method, and guide future measurements that could be carried out for example at the Atacama Large Millimeter/submillimeter Array (ALMA).

\section{Methods}

\subsection{Background}

The induced frequency shift of a probed transition is given as
\begin{equation}
\frac{\Delta\nu}{\nu_0}=T_{u,l}\frac{\Delta\mu}{\mu_0},
\label{eq.shift}
\end{equation}
where $\Delta\nu=\nu_{obs}-\nu_0$ is the change in the frequency and $\Delta\mu=\mu_{obs}-\mu_0$ is the change in $\mu$, both with respect to their accepted values $\nu_0$ and $\mu_0$. Using this relation one can easily show that the rotation-inversion transitions associated with the $\nu_2$ vibrational state of ammonia may exhibit induced frequency shifts more than one order of magnitude larger than the pure inversion transitions in the vibrational ground state, which are currently used in the probing of $\mu$ both temporally~\citep{Flambaum:2007,Murphy:2008,Menten:2008,Henkel:2009,Kanekar:2011} and spatially~\citep{Molaro:2009,Levshakov:2010b,Levshakov:2010a,Levshakov:2013}. Various $^{14}$NH$_3$ ro-inversional transitions have already been observed extraterrestrially~\citep{Mauersberger:1988,Schilke:1990,Schilke:1992}, whilst others with notable sensitivities possess Einstein coefficients comparable to those of the observed transitions. It is legitimate then to expect their eventual extragalactic detection, and when combined with their enhanced sensitivity, there is scope for a major improvement of the current ammonia analyses.

Another promising anomaly exhibited by the spectra of ammonia is caused by the
so-called ``giant'' $l$-type doubling, which leads to a ``reversal'' of the
inversion doublets in the $K=1$ levels in the $+l$ component of the $\nu_4$
states of $^{14}$NH$_3$ and $^{15}$NH$_3$. The inversion doublets are reversed
because for $K=1$, only one of the $A_1$ or $A_2$ sublevels is shifted by the
Coriolis interactions, and only the $A_2$ states have non-zero spin statistical
weights (see Fig. 1 and \citet{Spirko:1976}). So far these transitions have not
been detected extraterrestrially. This is to be expected since the physical
temperatures prevailing in the interstellar medium are too low to provide
significant population of the aforementioned states. However they could be
effectively populated by exoergic chemical formation processes, resulting in
the detection of highly excited states~\citep{Mills:2013,Lis:2014}.
Interestingly, the `highest energy' $(J,K)=(18,18)$ line of $^{14}$NH$_3$
observed towards the galactic centre star forming region Sgr B2, corresponds to
the state lying $3130{\,}$K above the ground vibrational
state~\citep*{Wilson:2006}.

\begin{figure}
\centering
\label{fig:theoref1}
% \vspace*{-40mm}
\includegraphics[width=0.7\columnwidth]{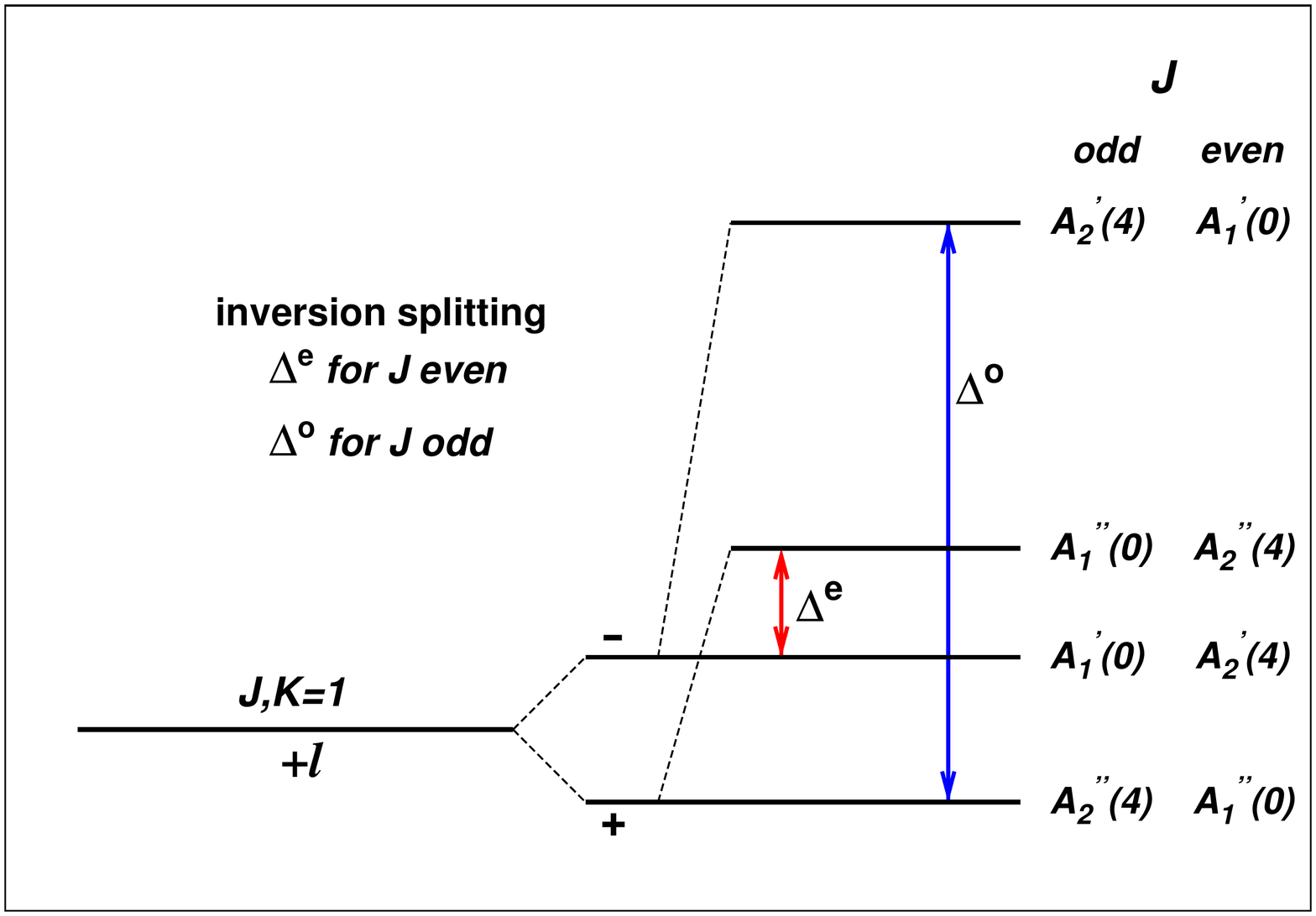}
% \vspace*{-8mm}
\caption{``Reversal'' of the inversion doublets in the +$l$ component of the $\nu_4$ level by the ``giant'' $l$-type doubling effect. Values in parentheses are the spin statistical weights.}
\end{figure}

The most common approach to computing sensitivity coefficients for a molecular system makes use of an effective Hamiltonian model, and determining how the parameters of this model depend on $\mu$~\citep{Jansen:2011a,Jansen:2011,Kozlov:2011,Levshakov:2011,Ilyushin:2012,Ilyushin:2014,Viatkina:2014}. For ammonia, the semiclassical Wentzel-Kramers-Brillouin (WKB) approximation has been used to obtain a general relationship to estimate the sensitivity of pure inversion frequencies in the ground vibrational state for $^{14}$NH$_3$~\citep{Flambaum:2007}, $^{15}$NH$_3$~\citep{Sartakov:2008}, $^{14}$ND$_3$~\citep{Flambaum:2007}, and $^{15}$ND$_3$~\citep{Veldhoven:2004}, whilst rotation-inversion transitions have been considered for the partly deuterated species $^{14}$NH$_2$D and $^{14}$ND$_2$H by \citet*{Kozlov:2010}.

The vibration-rotation-inversion transitions of $^{14}$NH$_3$ were investigated by \citet{Spirko:2014}, but theoretical calculations of the sensitivities using perturbation theory may not be entirely robust since the nominator and denominator in Eq.~(1) contain differences of large numbers. We thus find it worthwhile not only to check the literature data for $^{14}$NH$_3$ by means of highly accurate variational calculations, but to extend the treatment to $^{15}$NH$_3$, $^{14}$ND$_3$ and $^{15}$ND$_3$, which are equally valid probes of $\mu$. It is also straightforward to incorporate the so far unprobed $\nu_4$ states into the present study.

The advantage of our variational approach is that along with sensitivity coefficients, reliable theoretical transition frequencies can be generated if no experimental data is available, and for all selected transitions, Einstein $A$ coefficients can be calculated to guide future observations.

\subsection{Variational Calculations}

The variational nuclear motion program TROVE~\citep*{TROVE:2007} has provided highly accurate theoretical frequency, intensity, and thermodynamic data for both $^{14}$NH$_3$ and $^{15}$NH$_3$~\citep{09YuBaYa.NH3,10YaYuPa.NH3,11YuBaTe.method,Yurchenko:2011,14SoHYu.PH3,Yurchenko:2015}. We use the potential energy surface and computational setup as described in \citet{Yurchenko:2011} and \citet{Yurchenko:2015}, which can naturally be extended to treat $^{14}$ND$_3$ and $^{15}$ND$_3$. Here we only discuss the calculation of sensitivity coefficients, for the method used to compute transition frequencies and Einstein $A$ coefficients we refer the reader to \citet{09YuBaYa.NH3}.

We rely on the assumption that the baryonic matter may be treated equally~\citep{Dent:2007}, i.e. $\mu$ is assumed to be proportional to the molecular mass. It is then sufficient to perform a series of calculations employing suitably scaled values for the mass of ammonia. We choose the scaling coefficient $f_m=\lbrace0.9996,0.9998,1.0000,1.0002,1.0004\rbrace$ such that the scaled mass, $m_{\mathrm{NH_3}}^{\prime}=f_m \times m_{\mathrm{NH_3}}$. The mass dependency of any energy level can be found by using finite differences for (a) the $f_m=\lbrace0.9998,1.0002\rbrace$, and (b) the $f_m=\lbrace0.9996,1.0004\rbrace$ calculated energies. Both (a) and (b) should yield identical results, with the latter values used to verify the former. Numerical values for the derivatives ${\rm d}E/{\rm d}\mu$ are easily determined and then used in Eq.~(1), along with accurate experimental values for the transition frequencies, to calculate sensitivity coefficients. Calculations with $f_m=1.0000$ provide theoretical frequency data and Einstein $A$ coefficients.

The variational approach is powerful in that it allows a comprehensive treatment of a molecule to be undertaken. All possible transitions and their mass dependence can be calculated. This permits a simple exploration of the sensitivities for any molecule, provided the necessary steps have been taken to perform accurate variational calculations in the first place. As a cross-check, we also employ the nonrigid inverter theory~\citep{Spirko:1976,Spirko:1983} to compute sensitivity coefficients as was done by \citet{Spirko:2014}. In the following we evaluate both approaches. Note that the standard Herzberg convention~\citep{Herzberg:1945} is used to label the vibrational states of ammonia with the normal mode quantum numbers $v_1$, $v_2$, $v_3$, $v_4$, $l_3$ and $l_4$. The $\nu_2$ state corresponds to the singly excited inversion mode $v_2=1$, whilst $\nu_4$ is the singly excited asymmetric bending mode $v_4=|l_4|=1$ (see \citet{Down:2013} for more details).

\section{Results and Discussion}

The variationally calculated sensitivities for $^{14}$NH$_3$ and $^{15}$NH$_3$ are listed in Tables \ref{tab:v2_14nh3_15nh3_1} to \ref{tab:roinv_ground_14nh3_15nh3}. The results are consistent with previous perturbative values~\citep{Spirko:2014} obtained using the nonrigid inverter theory approach~\citep{Spirko:1976,Spirko:1983}, and `Born-Oppenheimer' estimates from \citet{Jansen:2014} (subsequently referred to as JBU). For transitions involving the $\nu_2$ vibrational states shown in Tables \ref{tab:v2_14nh3_15nh3_1}, \ref{tab:v2_14nh3_15nh3_2} and \ref{tab:v2_14nh3_15nh3_3}, the agreement is near quantitative with the exception of the ``forbidden'' combination difference $|a,J\!\!=\!\!3,K\!\!=\!\!3,v_{2}\!\!=\!\!1\rangle$ - $|s,J\!\!=\!\!3,K\!\!\!=\!\!0,v_{2}\!~\!=\!\!1\rangle$. The profoundly different sensitivities for these transitions when going from $^{14}$NH$_3$ to $^{15}$NH$_3$ is of particular interest. A possible variation of $\mu$ requires the measurement of at least two transitions with differing sensitivities. In the case of $|a,J\!\!=\!\!3,K\!\!=\!\!3,v_{2}\!\!=\!\!1\rangle$ - $|s,J\!\!=\!\!3,K\!\!\!=\!\!0,v_{2}\!~\!=\!\!1\rangle$, both isotopologues possess a large value of $T$. Importantly though they are of opposite sign, thus enabling a conclusive detection with regard to these particular transitions. An all-ammonia observation of a drifting $\mu$ would circumvent some of the intrinsic difficulties that appear when using other reference molecules~\citep{Murphy:2008,Henkel:2009,Kanekar:2011}, which may not necessarily reside at the same location in space.

\begin{table*}
\vspace*{-0.0cm}
\caption{The rotation-inversion frequencies ($\nu$), Einstein coefficients ($A$), and sensitivities ($T$) of $^{14}$NH$_3$ and their $^{15}$NH$_3$ counterparts in the $\nu_2$ vibrational state.}
\label{tab:v2_14nh3_15nh3_1}
%\begin{ruledtabular}
\resizebox{\linewidth}{!}{\begin{tabular}{c @{\extracolsep{0.01in}}
                c @{\extracolsep{0.01in}}
                c @{\extracolsep{0.01in}}
                c @{\extracolsep{0.01in}}
                c @{\extracolsep{0.1000in}}
                c @{\extracolsep{0.01in}}
                c @{\extracolsep{0.01in}}
                c @{\extracolsep{0.01in}}
                c @{\extracolsep{0.0100in}}
                c @{\extracolsep{0.2000in}}
                c @{\extracolsep{0.1500in}}
                c @{\extracolsep{0.1500in}}
                c}
%& & & & & & & &  \\
\hline \\[-2mm]
\multicolumn{1}{c}{$\Gamma\p$}&\multicolumn{1}{c}{$p\p$}&\multicolumn{1}{c}{$J\p$}&\multicolumn{1}{c}{$K\p$}&\multicolumn{1}{c}{$v_{2}\p$}&\multicolumn{1}{c}{$\Gamma\pp$}&
\multicolumn{1}{c}{$p\pp$}&\multicolumn{1}{c}{$J\pp$}&\multicolumn{1}{c}{$K\pp$}&\multicolumn{1}{c}{$v_{2}\pp$}&
\multicolumn{1}{c}{$\nu$/MHz}&\multicolumn{1}{c}{$A$/s$^{-1}$}&\multicolumn{1}{c}{$T$} \\[1mm]
\hline \\[-1mm]
  & &   &   &   &   &   & & $^{14}$NH$_3$     &  &         &                \\[1mm]
$E\pp$ & s & 2 & 1 & 1 & $E\p$ & a & 1 & 1 & 1 & 140142$^a$ & 0.1474$\times 10^{-4}$ & 17.24(16.92$^b$)  \\[0.3mm]
$A_2\pp$ & a & 0 & 0 & 1 & $A_2\p$ & s & 1 & 0 & 1 & 466244$^c$ & 0.1824$\times 10^{-2}$ & -6.587(-6.409)    \\[1mm]
  &  &   &   &   &   &   & & $^{15}$NH$_3$     &  &         &                \\[1mm]
$E\pp$ & s & 2 & 1 & 1 & $E\p$ & a & 1 & 1 & 1 & 175053     & 0.2939$\times 10^{-4}$ & 13.33(13.28)      \\[0.3mm]
$A_2\pp$ & a & 0 & 0 & 1 & $A_2\p$ & s & 1 & 0 & 1 & 430038     & 0.1425$\times 10^{-2}$ & -6.894(-6.908)    \\[1mm]
\hline \\[-3mm]
%\\
%\hline \\[-1mm]
\end{tabular}}
\vspace*{1mm}
\\
\footnotesize $^{14}$NH$_3$: Einstein coefficients from \citet{Yurchenko:2011}; $^a$Astronomical observation from \citet*{Mauersberger:1988} and \citet{Schilke:1990}; $^b$JBU sensitivity coefficient reaches a value of 18.8 (see \citet{Jansen:2014}); $^c$Astronomical observation from \citet{Schilke:1992}; values in parentheses from \citet{Spirko:2014}, obtained using the nonrigid inverter theory. \\
$^{15}$NH$_3$: Frequencies and Einstein coefficients from \citet{Urban:1985} and \citet{Yurchenko:2015}, respectively; values in parentheses obtained using the nonrigid inverter theory with the frequencies from \citet{Urban:1985}. \\
%\end{ruledtabular}
\end{table*}

\begin{table*}
\vspace*{-0.0cm}
\caption{The wavenumbers ($\nu$), wavelengths ($\lambda$), Einstein coefficients ($A$), and sensitivities ($T$) for transitions between the ground and $\nu_2$ vibrational state of $^{14}$NH$_3$ and their $^{15}$NH$_3$ counterparts.}
\label{tab:v2_14nh3_15nh3_2}
%\begin{ruledtabular}
\resizebox{\linewidth}{!}{\begin{tabular}{c @{\extracolsep{0.01in}}
                c @{\extracolsep{0.01in}}
                c @{\extracolsep{0.01in}}
                c @{\extracolsep{0.01in}}
                c @{\extracolsep{0.10in}}
                c @{\extracolsep{0.0100in}}
                c @{\extracolsep{0.01in}}
                c @{\extracolsep{0.01in}}
                c @{\extracolsep{0.0100in}}
                c @{\extracolsep{0.2000in}}
                c @{\extracolsep{0.2000in}}
                c @{\extracolsep{0.2200in}}
                c @{\extracolsep{0.2200in}}
                c}
%& & & & & & & &  \\
\hline \\[-2mm]
\multicolumn{1}{c}{$\Gamma\p$}&\multicolumn{1}{c}{$p\p$}&\multicolumn{1}{c}{$J\p$}&\multicolumn{1}{c}{$K\p$}&\multicolumn{1}{c}{$v_{2}\p$}&\multicolumn{1}{c}{$\Gamma\pp$}&
\multicolumn{1}{c}{$p\pp$}&\multicolumn{1}{c}{$J\pp$}&\multicolumn{1}{c}{$K\pp$}&\multicolumn{1}{c}{$v_{2}\pp$}&
\multicolumn{1}{c}{$\nu$/cm$^{-1}$}&\multicolumn{1}{c}{$\lambda$/$\mu$m}&\multicolumn{1}{c}{$A$/s$^{-1}$}&\multicolumn{1}{c}{$T$}\\[1mm]
\hline \\[-1mm]
  &  &   &   &   &   &  & & & $^{14}$NH$_3$     &  &         &                  \\[1mm]
$A_2\p$ & s & 6 & 6 & 1 & $A_2\pp$ & a & 6 & 6 & 0 & 927.3230 & 10.7837 & 0.1316$\times 10^{+2}$ & -0.367(-0.356) \\[0.3mm]
$E\p$ & s & 2 & 2 & 1 & $E\pp$ & a & 2 & 2 & 0 & 931.3333 & 10.7373 & 0.1030$\times 10^{+2}$ & -0.371(-0.366) \\[0.3mm]
$E\pp$ & s & 2 & 1 & 1 & $E\p$ & a & 1 & 1 & 0 & 971.8821 & 10.2893 & 0.5238$\times 10^{+1}$ & -0.399(-0.394) \\[0.3mm]
$E\pp$ & s & 1 & 1 & 1 & $E\p$ & a & 2 & 1 & 0 & 891.8820 & 11.2122 & 0.6795$\times 10^{+1}$ & -0.344(-0.339) \\[0.3mm]
$A_2\p$ & s & 1 & 0 & 1 & $A_2\pp$ & a & 2 & 0 & 0 & 892.1567 & 11.2088 & 0.9054$\times 10^{+1}$ & -0.344(-0.339) \\[0.3mm]
$A_2\pp$ & s & 3 & 3 & 1 & $A_2\p$ & a & 3 & 3 & 0 & 930.7571 & 10.7439 & 0.1158$\times 10^{+2}$ & -0.370(-0.366) \\[1mm]
  &  &   &   &   &   &  & & & $^{15}$NH$_3$     &  &         &                  \\[1mm]
$A_2\p$ & s & 6 & 6 & 1 & $A_2\pp$ & a & 6 & 6 & 0 & 923.4541 & 10.8289 & 0.1290$\times 10^{+2}$ & -0.365(-0.365) \\[0.3mm]
$E\p$ & s & 2 & 2 & 1 & $E\pp$ & a & 2 & 2 & 0 & 927.4034 & 10.7828 & 0.1010$\times 10^{+2}$ & -0.373(-0.373) \\[0.3mm]
$E\pp$ & s & 2 & 1 & 1 & $E\p$ & a & 1 & 1 & 0 & 967.8597 & 10.3321 & 0.5133$\times 10^{+1}$ & -0.400(-0.400) \\[0.3mm]
$E\pp$ & s & 1 & 1 & 1 & $E\p$ & a & 2 & 1 & 0 & 888.0413 & 11.2607 & 0.6664$\times 10^{+1}$ & -0.345(-0.345) \\[0.3mm]
$A_2\p$ & s & 1 & 0 & 1 & $A_2\pp$ & a & 2 & 0 & 0 & 888.3174 & 11.2572 & 0.8878$\times 10^{+1}$ & -0.346(-0.346) \\[0.3mm]
$A_2\pp$ & s & 3 & 3 & 1 & $A_2\p$ & a & 3 & 3 & 0 & 926.8378 & 10.7894 & 0.1135$\times 10^{+2}$ & -0.372(-0.372) \\[1mm]
\hline \\[-3mm]
\end{tabular}}
\vspace*{1mm}
\\
\footnotesize $^{14}$NH$_3$: Wavenumbers and Einstein coefficients from \citet{Urban:1984} and \citet{Yurchenko:2011}, respectively; Astronomical observations reported in \citet{Betz:1979} and \citet{Evans:1991}; values in parentheses from \citet{Spirko:2014}, obtained using the nonrigid inverter theory. \\
$^{15}$NH$_3$: Wavenumbers provided by Fusina, Di Lonardo \& Predoi-Cross (in preparation), Einstein coefficients from \citet{Yurchenko:2015}; values in parentheses obtained using the nonrigid inverter theory with the frequencies from Fusina, Di Lonardo \& Predoi-Cross (in preparation). \\
%\end{ruledtabular}
\end{table*}

%\newpage
\begin{table}
\vspace*{-1.5cm}
\caption{The vibration-rotation-inversion transitions associated with the $|a,J,K\!\!=\!\!3,v_{2}\!\!=\!\!1\rangle$ - $|s,J,K\!\!\!=\!\!0,v_{2}\!~\!=\!\!1\rangle$ resonances.}
\label{tab:v2_14nh3_15nh3_3}
%\begin{ruledtabular}
\resizebox{\linewidth}{!}{\begin{tabular}{c @{\extracolsep{0.01in}}
                c @{\extracolsep{0.0100in}}
                c @{\extracolsep{0.0100in}}
                c @{\extracolsep{0.0100in}}
                c @{\extracolsep{0.2000in}}
                c @{\extracolsep{0.0100in}}
                c @{\extracolsep{0.0100in}}
                c @{\extracolsep{0.0100in}}
                c @{\extracolsep{0.0100in}}
                c @{\extracolsep{0.3000in}}
                c @{\extracolsep{0.1200in}}
                c @{\extracolsep{0.1200in}}
                c @{\extracolsep{0.0800in}}
                c}
%& & & & & & & &  \\
\hline \\[-2mm]
\multicolumn{1}{c}{$\Gamma\p$}&\multicolumn{1}{c}{$p\p$}&\multicolumn{1}{c}{$J\p$}&\multicolumn{1}{c}{$K\p$}&\multicolumn{1}{c}{$v_{2}\p$}&\multicolumn{1}{c}{$\Gamma\pp$}&
\multicolumn{1}{c}{$p\pp$}&\multicolumn{1}{c}{$J\pp$}&\multicolumn{1}{c}{$K\pp$}&\multicolumn{1}{c}{$v_{2}\pp$}&
\multicolumn{1}{c}{$\nu$/MHz}&\multicolumn{1}{c}{$A$/s$^{-1}$}&
\multicolumn{1}{c}{$T$}&\multicolumn{1}{c}{Obs. Ref.} \\[2mm]
\hline \\[-1mm]
  &  &   &   &   &   &   & & $^{14}$NH$_3$     &  &         &                  \\[1mm]
$A_2\p$ & a & 3 & 3 & 1 & $A_2\pp$ & s & 3 & 3 & 0 & 29000313.7 & 0.1176$\times 10^{+2}$ & -0.484(-0.484)  & $^b$ \\[0.3mm]
$A_2\p$ & s & 3 & 0 & 1 & $A_2\pp$ & s & 3 & 3 & 0 & 28997430.0 & 0.2025$\times 10^{0}$ & -0.405(-0.405)  &  $^b$ \\[0.3mm]
 & a & 3 & 3 & 1 &  & s & 3 & 0 & 1 &     2883.7 &           & -790.6(-1001$^a$)&  $^b$ \\[1.4mm]
$A_2\p$ & a & 3 & 3 & 1 & $A_2\pp$ & a & 2 & 0 & 1 &   772594.9 & 0.6018$\times 10^{-4}$ & -0.868(-0.868)  & $^c$ \\[0.3mm]
$A_2\p$ & s & 3 & 0 & 1 & $A_2\pp$ & a & 2 & 0 & 1 &   769710.2 & 0.3471$\times 10^{-2}$ &  2.090(2.089)  &  $^c$ \\[0.3mm]
 & a & 3 & 3 & 1 &  & s & 3 & 0 & 1 &     2884.7 &           & -790.3(-1001)&  $^c$ \\[1.4mm]
$A_2\p$ & a & 3 & 3 & 1 & $A_2\pp$ & s & 3 & 3 & 1 &  1073050.7 & 0.1634$\times 10^{-1}$ & -3.350(-3.353)  & $^c$ \\[0.3mm]
$A_2\p$ & s & 3 & 0 & 1 & $A_2\pp$ & s & 3 & 3 & 1 &  1070166.6 & 0.2765$\times 10^{-3}$ & -1.228(-1.229)  & $^c$ \\[0.3mm]
 & a & 3 & 3 & 1 &  & s & 3 & 0 & 1 &     2884.1 &           & -790.5(-1001)& $^c$ \\[1.4mm]
$A_2\p$ & a & 5 & 3 & 1 & $A_2\pp$ & s & 5 & 3 & 0 & 28971340.5 & 0.4692$\times 10^{+1}$ & -0.484(-0.484)  & $^d$ \\[0.3mm]
$A_2\p$ & s & 5 & 0 & 1 & $A_2\pp$ & s & 5 & 3 & 0 & 29050552.5 & 0.2147$\times 10^{-2}$ & -0.408(-0.408)  & $^d$ \\[0.3mm]
 & a & 5 & 3 & 1 &  & s & 5 & 0 & 1 &    79212.0 &          &  27.38(27.35)  & $^d$ \\[1.4mm]
$A_2\p$ & a & 5 & 3 & 1 & $A_2\pp$ & s & 5 & 3 & 1 &   979649.1 & 0.5141$\times 10^{-2}$ & -3.425(-3.427)  & $^d$ \\[0.3mm]
$A_2\p$ & s & 5 & 0 & 1 & $A_2\pp$ & s & 5 & 3 & 1 &  1058861.1 & 0.3714$\times 10^{-5}$ &  -1.120(-1.120)  & $^d$ \\[0.3mm]
 & a & 5 & 3 & 1 &  & s & 5 & 0 & 1 &    79212.0 &          &  27.38(27.35)  & $^d$ \\[1.4mm]
$A_2\p$ & a & 5 & 3 & 1 & $A_2\pp$ & a & 4 & 0 & 1 &  1956241.1 & 0.4129$\times 10^{-4}$ & -0.988(-0.988)  & $^d$ \\[0.3mm]
$A_2\p$ & s & 5 & 0 & 1 & $A_2\pp$ & a & 4 & 0 & 1 &  2035453.1 & 0.7023$\times 10^{-1}$ &  0.116(0.116)  & $^d$ \\[0.3mm]
 & a & 5 & 3 & 1 &  & s & 5 & 0 & 1 &    79212.0 &           & 27.38(29.35) & $^d$ \\[1.4mm]
$A_2\p$ & a & 7 & 3 & 1 & $A_2\pp$ & s & 7 & 3 & 0 & 28934099.5 & 0.2399$\times 10^{+1}$ & -0.480(-0.480) & $^d$ \\[0.3mm]
$A_2\p$ & s & 7 & 0 & 1 & $A_2\pp$ & s & 7 & 3 & 0 & 29118808.5 & 0.1095$\times 10^{-3}$ & -0.416(-0.416) & $^d$ \\[0.3mm]
 & a & 7 & 3 & 1 &  & s & 7 & 0 & 1 &   184709.0 &          &  9.561(9.582) &  $^d$ \\[1.4mm]
$A_2\p$ & a & 9 & 3 & 1 & $A_2\pp$ & s & 9 & 3 & 0 & 28892089.9 & 0.1444$\times 10^{+1}$ & -0.475(-0.475) & $^d$ \\[0.3mm]
$A_2\p$ & s & 9 & 0 & 1 & $A_2\pp$ & s & 9 & 3 & 0 & 29194454.6 & 0.1029$\times 10^{-3}$ & -0.425(-0.425) & $^d$ \\[0.3mm]
 & a & 9 & 3 & 1 &  & s & 9 & 0 & 1 &   302364.7 &          &  4.350(4.363) & $^d$ \\[1mm]
  &  &   &   &   &   &   & & $^{15}$NH$_3$     &  &         &           &       \\[1mm]
$A_2\p$ & a & 3 & 3 & 1 & $A_2\pp$ & s & 3 & 3 & 0 & 28843885.0 & 0.1171$\times 10^{+2}$ & -0.486(-0.486) & $^e$ \\[0.3mm]
$A_2\p$ & s & 3 & 0 & 1 & $A_2\pp$ & s & 3 & 3 & 0 & 28872669.9 & 0.2187$\times 10^{-2}$ & -0.403(-0.403) & $^e$ \\[0.3mm]
 & a & 3 & 3 & 1 &  & s & 3 & 0 & 1 &    28784.9 &          &  82.96(81.69) & $^e$ \\[1.4mm]
$A_2\p$ & a & 3 & 3 & 1 & $A_2\pp$ & a & 2 & 0 & 1 &   774222.8 & 0.7160$\times 10^{-6}$ & -0.999(-0.999 ) & $^f$ \\[0.3mm]
$A_2\p$ & s & 3 & 0 & 1 & $A_2\pp$ & a & 2 & 0 & 1 &   802986.7 & 0.4035$\times 10^{-2}$ &  2.011(2.010) & $^f$ \\[0.3mm]
 & a & 3 & 3 & 1 &  & s & 3 & 0 & 1 &    28763.9 &          &  83.02(81.69) & $^f$ \\[1.4mm]
$A_2\p$ & a & 3 & 3 & 1 & $A_2\pp$ & s & 3 & 3 & 1 &  1035207.4 & 0.1491$\times 10^{-1}$ & -3.473(-3.476) & $^f$ \\[0.3mm]
$A_2\p$ & s & 3 & 0 & 1 & $A_2\pp$ & s & 3 & 3 & 1 &  1063971.3 & 0.3245$\times 10^{-5}$ & -1.228(-1.135) & $^f$ \\[0.3mm]
 & a & 3 & 3 & 1 &  & s & 3 & 0 & 1 &    28763.9 &          &  83.02(81.69) & $^f$ \\[1.4mm]
$A_2\p$ & a & 5 & 3 & 1 & $A_2\pp$ & s & 5 & 3 & 0 & 28817906.5 & 0.4598$\times 10^{+1}$ & -0.483(-0.483) & $^e$ \\[0.3mm]
$A_2\p$ & s & 5 & 0 & 1 & $A_2\pp$ & s & 5 & 3 & 0 & 28927141.3 & 0.7768$\times 10^{-3}$ & -0.409(-0.409) & $^e$ \\[0.3mm]
 & a & 5 & 3 & 1 &  & s & 5 & 0 & 1 &   109234.8 &          &  19.02(19.02) &  $^e$ \\[1.4mm]
$A_2\p$ & a & 5 & 3 & 1 & $A_2\pp$ & s & 5 & 3 & 1 &   943226.9 & 0.4588$\times 10^{-2}$ & -3.453(-3.455) & $^f$ \\[0.3mm]
$A_2\p$ & s & 5 & 0 & 1 & $A_2\pp$ & s & 5 & 3 & 1 &  1052459.7 & 0.1548$\times 10^{-5}$ & -1.120(-1.121) & $^f$ \\[0.3mm]
 & a & 5 & 3 & 1 &  & s & 5 & 0 & 1 &   109232.8 &          &  19.04(19.02) & $^f$ \\[1.4mm]
$A_2\p$ & a & 5 & 3 & 1 & $A_2\pp$ & a & 4 & 0 & 1 &  1955711.7 & 0.1882$\times 10^{-4}$ & -0.988(-0.988) & $^f$ \\[0.3mm]
$A_2\p$ & s & 5 & 0 & 1 & $A_2\pp$ & a & 4 & 0 & 1 &  2064944.5 & 0.7369$\times 10^{-1}$ &  0.071(0.071) & $^f$ \\[0.3mm]
 & a & 5 & 3 & 1 &  & s & 5 & 0 & 1 &   109232.8 &          &  19.05(19.02) & $^f$ \\[1.4mm]
$A_2\p$ & a & 7 & 3 & 1 & $A_2\pp$ & s & 7 & 3 & 0 & 28784706.6 & 0.2399$\times 10^{+1}$ & -0.479(-0.479) & $^e$ \\[0.3mm]
$A_2\p$ & s & 7 & 0 & 1 & $A_2\pp$ & s & 7 & 3 & 0 & 28997286.1 & 0.1095$\times 10^{-3}$ & -0.418(-0.418) & $^e$ \\[0.3mm]
 & a & 7 & 3 & 1 &  & s & 7 & 0 & 1 &   212579.5 &          &  7.898(7.073) & $^e$ \\[1.4mm]
$A_2\p$ & a & 9 & 3 & 1 & $A_2\pp$  & s & 9 & 3 & 0 & 28747714.9 & 0.1444$\times 10^{+1}$ & -0.479(-0.475) & $^e$ \\[0.3mm]
$A_2\p$ & s & 9 & 0 & 1 & $A_2\pp$ & s & 9 & 3 & 0 & 29075088.5 & 0.1029$\times 10^{-3}$ & -0.418(-0.427) & $^e$ \\[0.3mm]
 & a & 9 & 3 & 1 &  & s & 9 & 0 & 1 &   327373.6 &          &  3.782(3.782) & $^e$ \\[1mm]
%\\
\hline \\[-3mm]
\end{tabular}}
%\end{ruledtabular}
\vspace*{1mm}
\\
\footnotesize $^{14}$NH$_3$: Einstein coefficients from \citet{Yurchenko:2011}; $^a$JBU sensitivity coefficient reaches a value of -938 (see \citet{Jansen:2014}); values in parentheses obtained using the nonrigid inverter theory with the calculated TROVE frequencies; $^b$\citet{Fichoux}; $^c$\citet{Belov}; $^d$\citet{Urban:1984}. \\
$^{15}$NH$_3$: Einstein coefficients from \citet{Yurchenko:2015}; \hspace{8mm} values in parentheses obtained using the nonrigid inverter theory with the calculated TROVE frequencies; $^e$Fusina, Di Lonardo \& Predoi-Cross (in preparation); $^f$\citet{Urban:1985}.\\
\end{table}

\begin{table*}
\vspace*{-0.0cm}
\caption{Inversion frequencies ($\nu$), Einstein coefficients ($A$), and sensitivities ($T$) of $^{14}$NH$_3$ and their $^{15}$NH$_3$ counterparts in the ground vibrational state.}
\label{tab:inv_ground_14nh3_15nh3}
%\begin{ruledtabular}
\resizebox{\linewidth}{!}{\begin{tabular}{c @{\extracolsep{0.01in}}
                c @{\extracolsep{0.10in}}
                c @{\extracolsep{0.0800in}}
                c @{\extracolsep{0.0800in}}
                c @{\extracolsep{0.2500in}}
                c @{\extracolsep{0.01in}}
                c @{\extracolsep{0.0400in}}
                c @{\extracolsep{0.0800in}}
                c @{\extracolsep{0.0800in}}
                c @{\extracolsep{0.0500in}}
                c}
%& & & & & & & &  \\
\hline \\[-2mm]
\multicolumn{1}{c}{$J$}&\multicolumn{1}{c}{$K$}&\multicolumn{1}{c}{$\nu$/MHz}&
\multicolumn{1}{c}{$A$/s$^{-1}$}&\multicolumn{1}{c}{$T$}&
\multicolumn{1}{c}{$J$}&\multicolumn{1}{c}{$K$}&\multicolumn{1}{c}{$\nu$/MHz}&
\multicolumn{1}{c}{$A$/s$^{-1}$}&\multicolumn{1}{c}{$T$} \\[1mm]
\hline \\[-1mm]
 &  &  &  &  &    \hspace*{-12mm}$^{14}$NH$_3$ &   &  &  &        \\[1mm]
 1 & 1 & 23694.3 & 0.1657$\times 10^{-6}$ & -4.310(-4.365) & 4 & 3 & 22688.3 & 0.1311$\times 10^{-6}$ & -4.289(-4.514) \\
 2 & 1 & 23098.8 & 0.5123$\times 10^{-7}$ & -4.297(-4.413) & 4 & 4 & 24139.4 & 0.2797$\times 10^{-6}$ & -4.317(-4.471) \\
 2 & 2 & 23722.5 & 0.2216$\times 10^{-6}$ & -4.311(-4.385) & 5 & 1 & 19838.3 & 0.6540$\times 10^{-8}$ & -4.220(-4.700) \\
 3 & 2 & 22834.2 & 0.9902$\times 10^{-7}$ & -4.288(-4.464) & 5 & 2 & 20371.5 & 0.2828$\times 10^{-7}$ & -4.231(-4.546) \\
 3 & 3 & 23870.1 & 0.2538$\times 10^{-6}$ & -4.312(-4.419) & 5 & 3 & 21285.3 & 0.7239$\times 10^{-7}$ & -4.257(-4.634) \\
 4 & 1 & 21134.3 & 0.1182$\times 10^{-7}$ & -4.249(-4.568) & 5 & 4 & 22653.0 & 0.1546$\times 10^{-6}$ & -4.282(-4.592) \\
 4 & 2 & 21703.4 & 0.5114$\times 10^{-7}$ & -4.262(-4.545) & 5 & 5 & 24533.0 & 0.3053$\times 10^{-6}$ & -4.327(-4.509) \\[1mm]
&  &  &  &  &    \hspace*{-12mm}$^{15}$NH$_3$ &   &  &  &        \\[1mm]
 1 & 1 & 22624.9 & 0.1464$\times 10^{-6}$ & -4.352(-4.333) & 4 & 3 & 21637.9 & 0.1149$\times 10^{-6}$ & -4.330(-4.309) \\
 2 & 1 & 22044.2 & 0.4521$\times 10^{-7}$ & -4.341(-4.321) & 4 & 4 & 23046.0 & 0.2469$\times 10^{-6}$ & -4.360(-4.341) \\
 2 & 2 & 22649.8 & 0.1958$\times 10^{-6}$ & -4.349(-4.330) & 5 & 1 & 18871.5 & 0.5729$\times 10^{-8}$ & -4.264(-4.239) \\
 3 & 2 & 21783.9 & 0.8730$\times 10^{-7}$ & -4.333(-4.312) & 5 & 2 & 19387.4 & 0.2480$\times 10^{-7}$ & -4.278(-4.254) \\
 3 & 3 & 22789.4 & 0.2241$\times 10^{-6}$ & -4.356(-4.337) & 5 & 3 & 20272.1 & 0.6358$\times 10^{-7}$ & -4.299(-4.276) \\
 4 & 1 & 20131.4 & 0.1039$\times 10^{-7}$ & -4.293(-4.270) & 5 & 4 & 21597.9 & 0.1360$\times 10^{-6}$ & -4.330(-4.309) \\
 4 & 2 & 20682.8 & 0.4498$\times 10^{-7}$ & -4.306(-4.284) & 5 & 5 & 23422.0 & 0.2695$\times 10^{-6}$ & -4.366(-4.347) \\[1mm]
\hline \\[-3mm]
\end{tabular}}
\vspace*{1mm}
\\
\footnotesize $^{14}$NH$_3$: Frequencies and Einstein coefficients from \citet{Lovas:2009} and \citet{Yurchenko:2011}, respectively; values in parentheses from \citet{Spirko:2014}, obtained using the nonrigid inverter theory. \\
$^{15}$NH$_3$: Frequencies and Einstein coefficients from \citet{Urban:1985} and \citet{Yurchenko:2015}, respectively; values in parentheses obtained using the nonrigid inverter theory with the frequencies from \citet{Urban:1985}. \\
%\end{ruledtabular}
\end{table*}

\begin{table*}
\vspace*{-0.0cm}
\caption{The rotation-inversion frequencies ($\nu$), Einstein coefficients ($A$), and sensitivities ($T$) of $^{14}$NH$_3$ and their $^{15}$NH$_3$ counterparts in the ground vibrational state.}
\label{tab:roinv_ground_14nh3_15nh3}
%\begin{ruledtabular}
\resizebox{\linewidth}{!}{\begin{tabular}{c @{\extracolsep{0.01in}}
                c @{\extracolsep{0.01in}}
                c @{\extracolsep{0.01in}}
                c @{\extracolsep{0.01in}}
                c @{\extracolsep{0.1in}}
                c @{\extracolsep{0.01in}}
                c @{\extracolsep{0.01in}}
                c @{\extracolsep{0.01in}}
                c @{\extracolsep{0.01in}}
                c @{\extracolsep{0.2000in}}
                c @{\extracolsep{0.1500in}}
                c @{\extracolsep{0.1500in}}
                c @{\extracolsep{0.0500in}}
                c}
%& & & & & & & &  \\
\hline \\[-2mm]
\multicolumn{1}{c}{$\Gamma\p$}&\multicolumn{1}{c}{$p\p$}&\multicolumn{1}{c}{$J\p$}&\multicolumn{1}{c}{$K\p$}&\multicolumn{1}{c}{$v_{2}\p$}&\multicolumn{1}{c}{$\Gamma\pp$}&
\multicolumn{1}{c}{$p\pp$}&\multicolumn{1}{c}{$J\pp$}&\multicolumn{1}{c}{$K\pp$}&\multicolumn{1}{c}{$v_{2}\pp$}&
\multicolumn{1}{c}{$\nu$/MHz}&\multicolumn{1}{c}{$A$/s$^{-1}$}&\multicolumn{1}{c}{$T$}& \\[1mm]
\hline \\[-1mm]
  &  &   &   &   &   &   & &$^{14}$NH$_3$  &   &         &           &       \\[1mm]
$A_2\p$ & s & 1 & 0 & 0 & $A_2\pp$ & a & 0 & 0 & 0 &  572498 & 0.1561$\times 10^{-2}$ & -0.860(-0.862) \\[0.3mm]
$A_2\pp$ & a & 2 & 0 & 0 & $A_2\p$ & s & 1 & 0 & 0 & 1214859 & 0.1791$\times 10^{-1}$ & -1.060(-1.063) \\[0.3mm]
$E\p$ & a & 2 & 1 & 0 & $E\pp$ & s & 1 & 1 & 0 & 1215245 & 0.1344$\times 10^{-1}$ & -1.061(-1.064) \\[1mm]
  &  &   &   &   &   &   & &$^{15}$NH$_3$  &   &         &           &       \\[1mm]
$A_2\p$ & s & 1 & 0 & 0 & $A_2\pp$ & a & 0 & 0 & 0 &  572112 & 0.1557$\times 10^{-2}$ &  -0.865(-0.866) \\[0.3mm]
$A_2\pp$ & a & 2 & 0 & 0 & $A_2\p$ & s & 1 & 0 & 0 & 1210889 & 0.1774$\times 10^{-1}$ &  -1.058(-1.058) \\[0.3mm]
$E\p$ & a & 2 & 1 & 0 & $E\pp$ & s & 1 & 1 & 0 & 1211277 & 0.1331$\times 10^{-1}$ &  -1.059(-1.059) \\[1mm]
\hline \\[-3mm]
\end{tabular}}
\vspace*{1mm}
\\
\footnotesize $^{14}$NH$_3$: Frequencies and Einstein coefficients from \citet{Persson:2010} and \citet{Yurchenko:2011}, respectively; values given in parentheses from \citet{Spirko:2014}, obtained using the nonrigid inverter theory. \\
$^{15}$NH$_3$: Frequencies and Einstein coefficients from \citet{Urban:1985} and \citet{Yurchenko:2015}, respectively; values in parentheses obtained using the nonrigid inverter theory with the frequencies from \citet{Urban:1985}. \\
%\end{ruledtabular}
\end{table*}

The inversion frequencies in the ground vibrational state, Table \ref{tab:inv_ground_14nh3_15nh3}, have comparable sensitivities for both $^{14}$NH$_3$ and $^{15}$NH$_3$, and this also holds true for the ro-inversional transitions shown in Table \ref{tab:roinv_ground_14nh3_15nh3}, demonstrating the validity of $^{15}$NH$_3$ as a probe of $\mu$. The sensitivity coefficients of the $\nu_4$ transitions shown in Table \ref{tab:v4_14nh3_15nh3}, although promising, do not acquire the impressive magnitudes of their $\nu_2$ counterparts. However the appearance of positive and negative values could be of real use in constraining $\mu$.

\begin{table*}
\vspace*{-0.0cm}
\caption{Inversion frequencies ($\nu$), Einstein coefficients ($A$), and sensitivities ($T$) of $^{14}$NH$_3$ and $^{15}$NH$_3$ in the $\nu_4$ vibrational state.}
\label{tab:v4_14nh3_15nh3}
%\begin{ruledtabular}
\begin{tabular}{c @{\extracolsep{0.001in}}
                c @{\extracolsep{0.001in}}
                c @{\extracolsep{0.0051in}}
                c @{\extracolsep{0.0200in}}
                c @{\extracolsep{0.1200in}}
                c @{\extracolsep{0.20000in}}
                c @{\extracolsep{0.001in}}
                c @{\extracolsep{0.001in}}
                c @{\extracolsep{0.00510in}}
                c @{\extracolsep{0.0200in}}
                c @{\extracolsep{0.1200in}}
                c}
%& & & & & & & &  \\
\hline \\[-2mm]
\multicolumn{1}{c}{$J$}&\multicolumn{1}{c}{$K$}&\multicolumn{1}{l}{$l$}&\multicolumn{1}{c}{$\nu$/MHz}&
\multicolumn{1}{c}{$A$/s$^{-1}$}&\multicolumn{1}{c}{$T$}&
\multicolumn{1}{c}{$J$}&\multicolumn{1}{c}{$K$}&\multicolumn{1}{l}{$l$}&\multicolumn{1}{c}{$\nu$/MHz}&
\multicolumn{1}{c}{$A$/s$^{-1}$}&\multicolumn{1}{c}{$T$} \\[1mm]
\hline \\[-1mm]
 &  &  &  &  &  &  \hspace*{-12mm}$^{14}$NH$_3$ &   &  &  & &       \\[1mm]
  1 &  1 &  -1 &  32400 & 0.4243$\times 10^{-6}$ & -4.268 & 4 &   3 &   1 & 57132 & 0.1968$\times 10^{-5}$ &  1.561 \\
  1 &  1 &   1 &  57843 & 0.2411$\times 10^{-5}$ & -2.234 & 4 &   2 &  -1 & 47526 & 0.5467$\times 10^{-6}$ & -1.550 \\
  2 &  2 &  -1 &  32111 & 0.5514$\times 10^{-6}$ & -4.250 & 4 &   2 &   1 & 46515 & 0.4020$\times 10^{-6}$ & -0.247 \\
  2 &  2 &   1 &  40189 & 0.1056$\times 10^{-5}$ & -2.381 & 4 &   1 &  -1 & 57681 & 0.2548$\times 10^{-6}$ & -0.220 \\
  2 &  1 &  -1 &  36797 & 0.2085$\times 10^{-6}$ & -3.133 & 4 &   1 &   1 &145888$^a$& 0.3787$\times 10^{-5}$ & -0.962 \\
  2 &  1 &   1 &  20655 & 0.3743$\times 10^{-7}$ &  2.720 & 5 &   5 &  -1 & 32037 & 0.6848$\times 10^{-6}$ & -4.264 \\
  3 &  3 &  -1 &  31893 & 0.6081$\times 10^{-6}$ & -4.259 & 5 &   5 &   1 & 68699 & 0.6198$\times 10^{-5}$ &  4.672 \\
  3 &  3 &   1 &  46679 & 0.1863$\times 10^{-5}$ & -0.667 & 5 &   4 &  -1 & 39071 & 0.8020$\times 10^{-6}$ & -2.832 \\
  3 &  2 &  -1 &  37500 & 0.4424$\times 10^{-6}$ & -2.961 & 5 &   4 &   1 & 73534 & 0.4807$\times 10^{-5}$ &  4.480 \\
  3 &  2 &   1 &  44963 & 0.6906$\times 10^{-6}$ & -1.023 & 5 &   3 &  -1 & 48346 & 0.8610$\times 10^{-6}$ & -1.506 \\
  3 &  1 &  -1 &  44755 & 0.1908$\times 10^{-6}$ & -1.687 & 5 &   3 &   1 & 64906 & 0.1799$\times 10^{-5}$ &  3.044 \\
  3 &  1 &   1 &177783$^a$&0.1087$\times 10^{-4}$& -0.482 & 5 &   2 &  -1 & 58699 & 0.6967$\times 10^{-6}$ & -0.181 \\
  4 &  4 &  -1 &  31884 & 0.6482$\times 10^{-6}$ & -4.258 & 5 &   2 &   1 & 44876 & 0.2025$\times 10^{-6}$ &  0.239 \\
  4 &  4 &   1 &  55765 & 0.3325$\times 10^{-5}$ &  1.668 & 5 &   1 &   1 & 78141$^a$& 0.4324$\times 10^{-6}$ &  0.990 \\
  4 &  3 &  -1 &  38460 & 0.6451$\times 10^{-6}$ & -2.855 & 5 &   1 &  -1 &380542$^a$& 0.4015$\times 10^{-4}$ & -0.178 \\[1mm]
&  &  &  &  &  &   \hspace*{-12mm}$^{15}$NH$_3$ &   &  &  &  &      \\[1mm]
  1 &   1 &  -1 & 31108 & 0.3758$\times 10^{-6}$ & -4.291 & 4 &   3 &   1 & 51989 & 0.1501$\times 10^{-5}$ &  0.684 \\
  1 &   1 &   1 & 55582 & 0.2142$\times 10^{-5}$ & -2.410 & 4 &   2 &  -1 & 44599 & 0.4524$\times 10^{-6}$ & -1.765 \\
  2 &   2 &  -1 & 30825 & 0.4880$\times 10^{-6}$ & -4.271 & 4 &   2 &   1 & 43225 & 0.3278$\times 10^{-6}$ & -0.728 \\
  2 &   2 &   1 & 37900 & 0.8883$\times 10^{-6}$ & -2.722 & 4 &   1 &  -1 & 53406 & 0.2029$\times 10^{-6}$ & -0.558 \\
  2 &   1 &  -1 & 34950 & 0.1788$\times 10^{-6}$ & -3.273 & 4 &   1 &   1 &146961$^a$ & 0.3870$\times 10^{-5}$ & -0.983 \\
  2 &   1 &   1 & 21904 & 0.4450$\times 10^{-7}$ &  2.351 & 5 &   5 &  -1 & 30732 & 0.6050$\times 10^{-6}$ & -4.280 \\
  3 &   3 &  -1 & 30606 & 0.5377$\times 10^{-6}$ & -4.281 & 5 &   5 &   1 & 61128$^a$ & 0.4341$\times 10^{-5}$ &  2.771 \\
  3 &   3 &   1 & 43275 & 0.1492$\times 10^{-5}$ & -1.358 & 5 &   4 &  -1 & 37071 & 0.6856$\times 10^{-6}$ & -2.937 \\
  3 &   2 &  -1 & 35551 & 0.3772$\times 10^{-6}$ & -3.082 & 5 &   4 &   1 & 65945$^a$ & 0.3431$\times 10^{-5}$ &  3.150 \\
  3 &   2 &   1 & 41928 & 0.5649$\times 10^{-6}$ & -1.494 & 5 &   3 &  -1 & 45373 & 0.7129$\times 10^{-6}$ & -1.689 \\
  3 &   1 &  -1 & 41947 & 0.1574$\times 10^{-6}$ & -1.941 & 5 &   3 &   1 & 59236$^a$ & 0.1351$\times 10^{-5}$ &  2.151 \\
  3 &   1 &   1&171460$^a$&0.9842$\times 10^{-5}$& -0.731 & 5 &   2 &  -1 & 54322 & 0.5536$\times 10^{-6}$ & -0.440 \\
  4 &   4 &  -1 & 30591 & 0.5729$\times 10^{-6}$ & -4.281 & 5 &   2 &   1 & 42037$^a$ & 0.1659$\times 10^{-6}$ & -0.242 \\
  4 &   4 &   1 & 50530 & 0.2502$\times 10^{-5}$ &  0.452 & 5 &   1 &  -1 & 71752$^a$ & 0.3362$\times 10^{-6}$ &  0.639 \\
  4 &   3 &  -1 & 36472 & 0.5506$\times 10^{-6}$ & -2.978 & 5 &   1 &   1 &369287$^a$ & 0.3728$\times 10^{-4}$ & -0.379 \\[1mm]
\hline \\[-3mm]
\end{tabular}
\vspace*{1mm}
\\
\footnotesize Frequencies from \citet{Cohen:1974} and \citet{Cohen:1980}; $^a$TROVE calculated value \\
%\end{ruledtabular}
\end{table*}

Because of the substantial differences in size of the inversion splittings, the mass sensitivity of the $^{14}$ND$_3$ and $^{15}$ND$_3$ transitions exhibit centrifugal distortion and Coriolis interaction dependence significantly different from that exhibited by $^{14}$NH$_3$ and $^{15}$NH$_3$ (see Tables \ref{tab:inv_ground_14nd3_15nd3}, \ref{tab:roinv_ground_14nd3_15nd3}, \ref{tab:v2_14nd3_15nd3_1}, \ref{tab:v2_14nd3_15nd3_2} and Fig. 2). The effects of these interactions are non negligible and must be included in any critical analysis. As only a small fraction of the total presence of ammonia in the interstellar medium is heavy ammonia, a detection of `higher energy' transitions is rather improbable. All the ammonia isotopomers appear as suitable targets for terrestrial studies however, such as those reported by \citet{Veldhoven:2004}, \citet{Sartakov:2008}, and \citet{Quintero:2014}.

\begin{figure}
\label{fig:theoref2}
%\vspace*{-5mm}
%\includegraphics[width=1.0\columnwidth]{one6.ps}
\hspace*{-7mm}\includegraphics[width=0.56\columnwidth,angle=0 ]{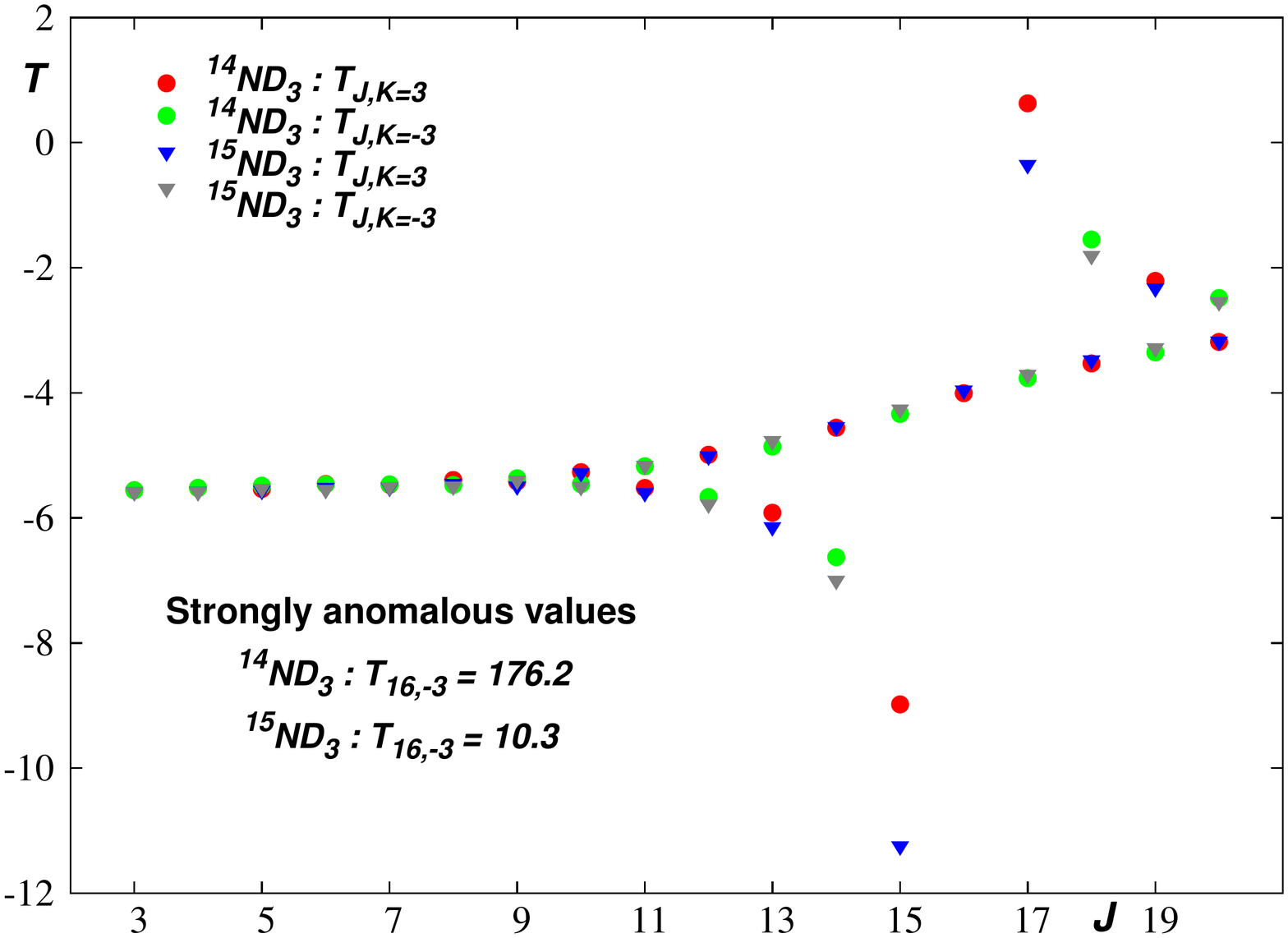}
% \vspace*{35mm}
\hspace*{-15mm}\includegraphics[width=0.56\columnwidth,angle=0]{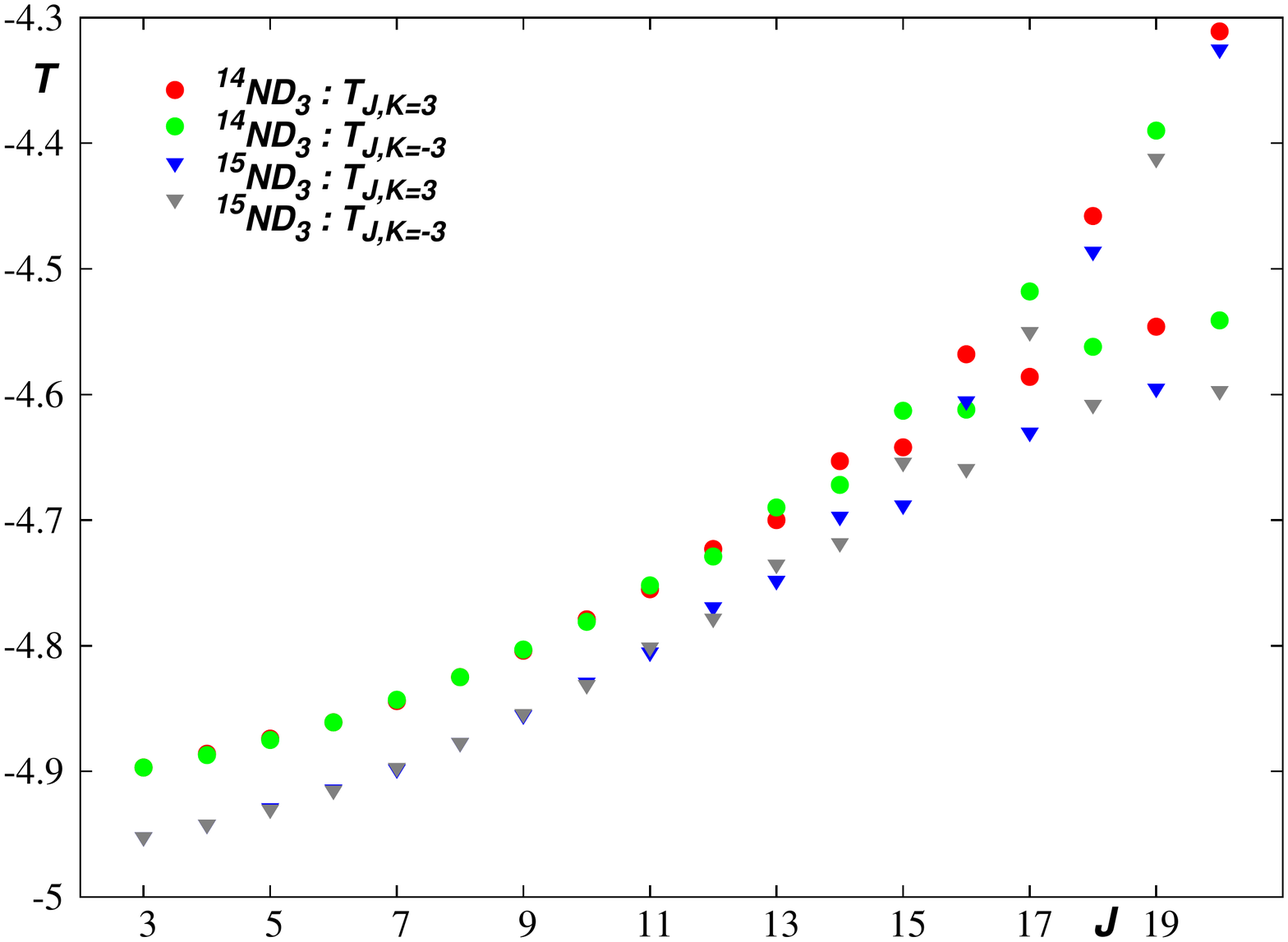}
% \vspace*{-48mm}
\caption{The sensitivities, $T$, of the inversion transitions of the $(J,K=\pm3)$ rotational states of $^{14}$ND$_3$ and $^{15}$ND$_3$ in the ground (left panel) and $\nu_2$ (right panel) vibrational states.}
\end{figure}

\begin{table*}
\vspace*{-0.0cm}
\caption{Inversion frequencies ($\nu$), Einstein coefficients ($A$), and sensitivities ($T$) of $^{14}$ND$_3$ and $^{15}$ND$_3$ in the ground vibrational state.}
\label{tab:inv_ground_14nd3_15nd3}
%\begin{ruledtabular}
\resizebox{\linewidth}{!}{\begin{tabular}{c @{\extracolsep{0.01in}}
                c @{\extracolsep{0.10in}}
                c @{\extracolsep{0.100in}}
                c @{\extracolsep{0.100in}}
                c @{\extracolsep{0.2500in}}
                c @{\extracolsep{0.01in}}
                c @{\extracolsep{0.0400in}}
                c @{\extracolsep{0.1000in}}
                c @{\extracolsep{0.1000in}}
                c @{\extracolsep{0.0500in}}
                c}
%& & & & & & & &  \\
\hline \\[-2mm]
\multicolumn{1}{c}{$J$}&\multicolumn{1}{c}{$K$}&\multicolumn{1}{c}{$\nu$/MHz}&
\multicolumn{1}{c}{$A$/s$^{-1}$}&\multicolumn{1}{c}{$T$}&
\multicolumn{1}{c}{$J$}&\multicolumn{1}{c}{$K$}&\multicolumn{1}{c}{$\nu$/MHz}&
\multicolumn{1}{c}{$A$/s$^{-1}$}&\multicolumn{1}{c}{$T$} \\[1mm]
\hline \\[-1mm]
 &  &  &  &  &    \hspace*{-12mm}$^{14}$ND$_3$ &   &  &  &        \\[1mm]
 1 &  1 &  1589.006 & 0.5764$\times 10^{-10}$& -5.541(-5.528)& 4 &  3 &  1558.600 & 0.4897$\times 10^{-10}$& -5.533(-5.520) \\
 2 &  1 &  1568.357 & 0.1849$\times 10^{-10}$& -5.556(-5.542)& 4 & -3 &  1558.178 & 0.4893$\times 10^{-10}$& -5.534(-5.521) \\
 2 &  2 &  1591.695 & 0.7721$\times 10^{-10}$& -5.543(-5.530)& 4 &  4 &  1612.997 & 0.9623$\times 10^{-10}$& -5.536(-5.525) \\
 3 &  1 &  1537.915 & 0.8725$\times 10^{-11}$& -5.526(-5.511)& 5 &  1 &  1450.435$^a$ & 0.2937$\times 10^{-11}$& -5.511(-5.493) \\
 3 &  2 &  1560.774 & 0.3644$\times 10^{-10}$& -5.537(-5.523)& 5 &  2 &  1471.785 & 0.1226$\times 10^{-10}$& -5.504(-5.487) \\
 3 &  3 &  1599.645 & 0.8810$\times 10^{-10}$& -5.571(-5.559)& 5 &  3 &  1507.525 & 0.2960$\times 10^{-10}$& -5.553(-5.537) \\
 3 & -3 &  1599.704 & 0.8811$\times 10^{-10}$& -5.571(-5.559)& 5 & -3 &  1509.218 & 0.2969$\times 10^{-10}$& -5.499(-5.484) \\
 4 &  1 &  1498.270 & 0.4848$\times 10^{-11}$& -5.503(-5.487)& 5 &  4 &  1561.146 & 0.5827$\times 10^{-10}$& -5.524(-5.511) \\
 4 &  2 &  1520.537 & 0.2025$\times 10^{-10}$& -5.493(-5.478)& 5 &  5 &  1631.784 & 0.1036$\times 10^{-9}$& -5.561(-5.551) \\[1mm]
&  &  &  &  &    \hspace*{-12mm}$^{15}$ND$_3$ &   &  &  &        \\[1mm]
 1 &  1 &  1430.340 & 0.4227$\times 10^{-10}$& -5.600(-5.577) & 4 &  3 &  1401.312 & 0.3578$\times 10^{-10}$& -5.600(-5.577) \\
 2 &  1 &  1410.980 & 0.1354$\times 10^{-10}$& -5.613(-5.589) & 4 & -3 &  1400.878 & 0.3575$\times 10^{-10}$& -5.602(-5.578) \\
 2 &  2 &  1432.641 & 0.5661$\times 10^{-10}$& -5.604(-5.581) & 4 &  2 &  1366.027 & 0.1476$\times 10^{-10}$& -5.586(-5.561) \\
 3 &  1 &  1382.510 & 0.5374$\times 10^{-11}$& -5.585(-5.560) & 5 &  1 &  1300.841$^a$ & 0.2130$\times 10^{-11}$& -5.562(-5.534) \\
 3 &  2 &  1403.684 & 0.2665$\times 10^{-10}$& -5.566(-5.542) & 5 &  2 &  1320.460$^a$ & 0.8907$\times 10^{-11}$& -5.575(-5.547) \\
 3 &  3 &  1439.719 & 0.5458$\times 10^{-10}$& -5.601(-5.579) & 5 &  3 &  1353.451 & 0.2153$\times 10^{-10}$& -5.585(-5.559) \\
 3 & -3 &  1439.783 & 0.6459$\times 10^{-10}$& -5.601(-5.579) & 5 & -3 &  1355.161 & 0.2162$\times 10^{-10}$& -5.551(-5.526) \\
 4 &  1 &  1345.533$^a$ & 0.3530$\times 10^{-11}$&  -5.564(-5.538) & 5 &  4 &  1403.179 & 0.4254$\times 10^{-10}$& -5.606(-5.583) \\
 4 &  2 &  1366.027 & 0.1476$\times 10^{-10}$& -5.586(-5.561) & 5 &  5 &  1468.666 & 0.7595$\times 10^{-10}$& -5.639(-5.619) \\[1mm]
\hline \\[-3mm]
\end{tabular}}
\vspace*{1mm}
\\
\footnotesize Unless stated otherwise, $^{14}$ND$_3$ and $^{15}$ND$_3$ frequencies from \citet{Murzin} and \citet{Carlotti}, respectively; values in parentheses obtained using the nonrigid inverter theory with the calculated TROVE frequencies; the $K=-3$ values refer to transitions between levels with spin statistical weight $=10$ ($A_{1}\p$, $A_{1}\pp$ species), the $K=3$ values refer to transitions between levels with spin statistical weight $=1$ ($A_{2}\p$, $A_{2}\pp$) species); $^a$\citet{Bester}.\\
\end{table*}

\begin{table*}
\vspace*{-0.0cm}
\caption{The rotation-inversion frequencies ($\nu$), Einstein coefficients ($A$), and sensitivities ($T$) of $^{14}$ND$_3$ and $^{15}$ND$_3$ in the ground vibrational state.}
\label{tab:roinv_ground_14nd3_15nd3}
%\begin{ruledtabular}
\begin{tabular}{c @{\extracolsep{0.01in}}
                c @{\extracolsep{0.01in}}
                c @{\extracolsep{0.01in}}
                c @{\extracolsep{0.01in}}
                c @{\extracolsep{0.100in}}
                c @{\extracolsep{0.01in}}
                c @{\extracolsep{0.01in}}
                c @{\extracolsep{0.0100in}}
                c @{\extracolsep{0.0100in}}
                c @{\extracolsep{0.200in}}
                c @{\extracolsep{0.1500in}}
                c @{\extracolsep{0.1500in}}
                c @{\extracolsep{0.0500in}}
                c}
%& & & & & & & &  \\
\hline \\[-2mm]
\multicolumn{1}{c}{$\Gamma\p$}&\multicolumn{1}{c}{$p\p$}&\multicolumn{1}{c}{$J\p$}&\multicolumn{1}{c}{$K\p$}&\multicolumn{1}{c}{$v_{2}\p$}&\multicolumn{1}{c}{$\Gamma\pp$}&
\multicolumn{1}{c}{$p\pp$}&\multicolumn{1}{c}{$J\pp$}&\multicolumn{1}{c}{$K\pp$}&\multicolumn{1}{c}{$v_{2}\pp$}&
\multicolumn{1}{c}{$\nu$/MHz}&\multicolumn{1}{c}{$A$/s$^{-1}$}&\multicolumn{1}{c}{$T$}& \\[1mm]
\hline \\[-1mm]
  &  &   &   &   &   &   & &$^{14}$ND$_3$  &   &         &           &       \\[1mm]
$A_1\pp$ & a & 1 & 0 & 0 & $A_1\p$ & s & 0 & 0 & 0 & 309909$^a$ & 0.2530$\times 10^{-3}$ & -1.022 \\[0.3mm]
$A_2\pp$ & a & 2 & 0 & 0 & $A_2\p$ & s & 1 & 0 & 0 & 618075$^a$ & 0.2409$\times 10^{-2}$ & -1.009 \\[0.3mm]
$E\p$ & a & 2 & 1 & 0 & $E\pp$ & s & 1 & 1 & 0 & 618124$^a$ & 0.1807$\times 10^{-2}$ & -1.009 \\[0.3mm]
$A_2\p$ & s & 1 & 0 & 0 & $A_2\pp$ & a & 0 & 0 & 0 & 306737$^a$ & 0.2450$\times 10^{-3}$ & -0.973 \\[0.3mm]
$A_1\p$ & s & 2 & 0 & 0 & $A_1\pp$ & a & 1 & 0 & 0 & 614933$^a$ & 0.2371$\times 10^{-2}$ & -0.985 \\[0.3mm]
$E\pp$ & s & 2 & 1 & 0 & $E\p$ & a & 1 & 1 & 0 & 614968$^a$ & 0.1778$\times 10^{-2}$ & -0.985 \\[0.3mm]
$A_1\pp$ & a & 3 & 0 & 0 & $A_1\p$ & s & 2 & 0 & 0 & 925947 & 0.8681$\times 10^{-2}$ & -1.005 \\[0.3mm]
$E\p$ & a & 3 & 1 & 0 & $E\pp$ & s & 2 & 1 & 0 & 926018 & 0.7717$\times 10^{-2}$ & -1.005 \\[0.3mm]
$E\pp$ & a & 3 & 2 & 0 & $E\p$ & s & 2 & 2 & 0 & 926228 & 0.4824$\times 10^{-2}$ & -1.005 \\[0.3mm]
$A_2\p$ & s & 3 & 0 & 0 & $A_2\pp$ & a & 2 & 0 & 0 & 922857 & 0.8591$\times 10^{-2}$ &  -0.989 \\[0.3mm]
$E\pp$ & s & 3 & 1 & 0 & $E\p$ & a & 2 & 1 & 0 & 922911 & 0.7637$\times 10^{-2}$ &  -0.989 \\[0.3mm]
$E\p$ & s & 3 & 2 & 0 & $E\pp$ & a & 2 & 2 & 0 & 923076 & 0.4773$\times 10^{-2}$ &  -0.999 \\[1mm]
  &  &   &   &   &   &   & &$^{15}$ND$_3$  &   &         &           &       \\[1mm]
$A_1\pp$ & a & 1 & 0 & 0 & $A_1\p$ & s & 0 & 0 & 0 & 308606$^a$ & 0.2499$\times 10^{-3}$ & -1.020 \\[0.3mm]
$A_2\pp$ & a & 2 & 0 & 0 & $A_2\p$ & s & 1 & 0 & 0 & 615628$^a$ & 0.2381$\times 10^{-2}$ & -1.008 \\[0.3mm]
$E\p$ & a & 2 & 1 & 0 & $E\pp$ & s & 1 & 1 & 0 & 615677$^a$ & 0.1785$\times 10^{-2}$ & -1.009 \\[0.3mm]
$A_2\p$ & s & 1 & 0 & 0 & $A_2\pp$ & a & 0 & 0 & 0 & 305750$^a$ & 0.2427$\times 10^{-3}$ &  -0.975 \\[0.3mm]
$A_1\p$ & s & 2 & 0 & 0 & $A_1\pp$ & a & 1 & 0 & 0 & 612801$^a$ & 0.2346$\times 10^{-2}$ &  -0.987 \\[0.3mm]
$E\pp$ & s & 2 & 1 & 0 & $E\p$ & a & 1 & 1 & 0 & 612836$^a$ & 0.1760$\times 10^{-2}$ &  -0.987 \\[0.3mm]
$A_1\pp$ & a & 3 & 0 & 0 & $A_1\p$ & s & 2 & 0 & 0 & 922356& 0.8582$\times 10^{-2}$& -1.004 \\[0.3mm]
$E\p$ & a & 3 & 1 & 0 & $E\pp$ & s & 2 & 1 & 0 & 922426& 0.7628$\times 10^{-2}$& -1.004 \\[0.3mm]
$E\pp$ & a & 3 & 2 & 0 & $E\p$ & s & 2 & 2 & 0 & 922636& 0.4768$\times 10^{-2}$& -1.004 \\[0.3mm]
$A_2\p$ & s & 3 & 0 & 0 & $A_2\pp$ & a & 2 & 0 & 0 & 919577& 0.8501$\times 10^{-2}$&  -0.990 \\[0.3mm]
$E\pp$ & s & 3 & 1 & 0 & $E\p$ & a & 2 & 1 & 0 & 919632& 0.7556$\times 10^{-2}$&  -0.990 \\[0.3mm]
$E\p$ & s & 3 & 2 & 0 & $E\pp$ & a & 2 & 2 & 0 & 919800& 0.4723$\times 10^{-2}$&  -0.990 \\[1mm]
\hline \\[-3mm]
%\hline \\[-1mm]
\end{tabular}
\vspace*{1mm}
\\
\footnotesize Unless stated otherwise, frequencies from \cite{Bester}; $^a$\citet{Helminger1} and \citet{Helminger2}.\\
\end{table*}

\begin{table*}
\vspace*{-0.0cm}
\caption{The rotation-inversion frequencies ($\nu$), Einstein coefficients ($A$), and sensitivities ($T$) of $^{14}$ND$_3$ and $^{15}$ND$_3$ in the $\nu_2$ vibrational state.}
\label{tab:v2_14nd3_15nd3_1}
%\begin{ruledtabular}
\begin{tabular}{c @{\extracolsep{0.01in}}
                c @{\extracolsep{0.01in}}
                c @{\extracolsep{0.01in}}
                c @{\extracolsep{0.01in}}
                c @{\extracolsep{0.1in}}
                c @{\extracolsep{0.01in}}
                c @{\extracolsep{0.01in}}
                c @{\extracolsep{0.01in}}
                c @{\extracolsep{0.01in}}
                c @{\extracolsep{0.2000in}}
                c @{\extracolsep{0.1500in}}
                c @{\extracolsep{0.1500in}}
                c @{\extracolsep{0.0500in}}
                c}
%& & & & & & & &  \\
\hline \\[-2mm]
\multicolumn{1}{c}{$\Gamma\p$}&\multicolumn{1}{c}{$p\p$}&\multicolumn{1}{c}{$J\p$}&\multicolumn{1}{c}{$K\p$}&\multicolumn{1}{c}{$v_{2}\p$}&\multicolumn{1}{c}{$\Gamma\pp$}&
\multicolumn{1}{c}{$p\pp$}&\multicolumn{1}{c}{$J\pp$}&\multicolumn{1}{c}{$K\pp$}&\multicolumn{1}{c}{$v_{2}\pp$}&
\multicolumn{1}{c}{$\nu$/MHz}&\multicolumn{1}{c}{$A$/s$^{-1}$}&\multicolumn{1}{c}{$T$}& \\[1mm]
\hline \\[-1mm]
 &  &   &   &   &   &   & &$^{14}$ND$_3$  &   &         &           &       \\[1mm]
$A_1\pp$ & a & 1 & 0 & 1 & $A_1\p$ & s & 0 & 0 & 1 &  412847 & 0.4983$\times 10^{-3}$ & -2.030 \\[0.3mm]
$A_2\pp$ & a & 2 & 0 & 1 & $A_2\p$ & s & 1 & 0 & 1 &  718585 & 0.3131$\times 10^{-2}$ & -1.585 \\[0.3mm]
$E\p$ & a & 2 & 1 & 1 & $E\pp$ & s & 1 & 1 & 1 &  719092 & 0.2352$\times 10^{-2}$ & -1.588 \\[0.3mm]
$A_2\p$ & s & 1 & 0 & 1 & $A_2\pp$ & a & 0 & 0 & 1 &  200763 & 0.5423$\times 10^{-4}$ &  1.119 \\[0.3mm]
$A_1\p$ & s & 2 & 0 & 1 & $A_1\pp$ & a & 1 & 0 & 1 &  508364 & 0.1082$\times 10^{-2}$ & -0.170 \\[0.3mm]
$E\pp$ & s & 2 & 1 & 1 & $E\p$ & a & 1 & 1 & 1 &  507940 & 0.8088$\times 10^{-3}$ & -0.166 \\[0.3mm]
$A_1\pp$ & a & 3 & 0 & 1 & $A_1\p$ & s & 2 & 0 & 1 &  1023449 & 0.9673$\times 10^{-2}$ & -1.404 \\[0.3mm]
$E\p$ & a & 3 & 1 & 1 & $E\pp$ & s & 2 & 1 & 1 & 1023971 & 0.8608$\times 10^{-2}$ & -1.405 \\[0.3mm]
$E\pp$ & a & 3 & 2 & 1 & $E\p$ & s & 2 & 2 & 1 & 1025546 & 0.5399$\times 10^{-2}$ & -1.411 \\[0.3mm]
$A_2\p$ & s & 3 & 0 & 1 & $A_2\pp$ & a & 2 & 0 & 1 &  816294 & 0.4830$\times 10^{-2}$ & -0.491 \\[0.3mm]
$E\pp$ & s & 3 & 1 & 1 & $E\p$ & a & 2 & 1 & 1 &  815898 & 0.4286$\times 10^{-2}$ & -0.488 \\[0.3mm]
$E\p$ & s & 3 & 2 & 1 & $E\pp$ & a & 2 & 2 & 1 &  814696 & 0.2663$\times 10^{-2}$ & -0.480 \\[0.3mm]
$A_2\pp$ & a & 4 & 0 & 1 & $A_2\p$ & s & 3 & 0 & 1 & 1327334 & 0.2188$\times 10^{-1}$ & -1.304 \\[0.3mm]
$E\p$ & a & 4 & 1 & 1 & $E\pp$ & s & 3 & 1 & 1 & 1327865 & 0.2053$\times 10^{-1}$ & -1.305 \\[0.3mm]
$E\pp$ & a & 4 & 2 & 1 & $E\p$ & s & 3 & 2 & 1 & 1329473 & 0.1647$\times 10^{-1}$ & -1.309 \\[0.3mm]
$A_2\p$ & a & 4 & 3 & 1 & $A_2\pp$ & s & 3 & 3 & 1 & 1332194 & 0.9646$\times 10^{-2}$ & -1.317 \\[0.3mm]
$A_1\p$ & a & 4 &-3 & 1 & $A_1\pp$ & s & 3 &-3 & 1 & 1332194 & 0.9646$\times 10^{-2}$ & -1.317 \\[0.3mm]
$A_1\p$ & s & 4 & 0 & 1 & $A_1\pp$ & a & 3 & 0 & 1 & 1124392 & 0.1315$\times 10^{-1}$ & -0.637 \\[0.3mm]
$E\pp$ & s & 4 & 1 & 1 & $E\p$ & a & 3 & 1 & 1 & 1124025 & 0.1231$\times 10^{-1}$ & -0.636 \\[0.3mm]
$E\p$ & s & 4 & 2 & 1 & $E\pp$ & a & 3 & 2 & 1 & 1122914 & 0.9805$\times 10^{-2}$ & -0.630 \\[0.3mm]
$A_2\pp$ & s & 4 & 3 & 1 & $A_2\p$ & a & 3 & 3 & 1 & 1121023 & 0.5679$\times 10^{-2}$ & -0.621 \\[0.3mm]
$A_1\pp$ & s & 4 &-3 & 1 & $A_1\p$ & a & 3 &-3 & 1 & 1121023 & 0.5679$\times 10^{-2}$ & -0.621 \\[0.3mm]
$A_1\pp$ & a & 5 & 0 & 1 & $A_1\p$ & s & 4 & 0 & 1 & 1630141 & 0.4149$\times 10^{-1}$ & -1.239 \\[0.3mm]
$E\p$ & a & 5 & 1 & 1 & $E\pp$ & s & 4 & 1 & 1 & 1630681 & 0.3986$\times 10^{-1}$ & -1.240 \\[0.3mm]
$E\pp$ & a & 5 & 2 & 1 & $E\p$ & s & 4 & 2 & 1 & 1632314 & 0.3494$\times 10^{-1}$ & -1.243 \\[0.3mm]
$A_2\p$ & a & 5 & 3 & 1 & $A_2\pp$ & s & 4 & 3 & 1 & 1635074 & 0.2671$\times 10^{-1}$ & -1.249 \\[0.3mm]
$A_1\p$ & a & 5 &-3 & 1 & $A_1\pp$ & s & 4 &-3 & 1 & 1635075 & 0.2671$\times 10^{-1}$ & -1.249 \\[0.3mm]
$E\pp$ & a & 5 & 4 & 1 & $E\p$ & s & 4 & 4 & 1 & 1639027 & 0.1509$\times 10^{-1}$ & -1.258 \\[0.3mm]
$A_2\p$ & s & 5 & 0 & 1 & $A_2\pp$ & a & 4 & 0 & 1 & 1432485 & 0.2790$\times 10^{-1}$ & -0.722 \\[0.3mm]
$E\pp$ & s & 5 & 1 & 1 & $E\p$ & a & 4 & 1 & 1 & 1432151 & 0.2676$\times 10^{-1}$ & -0.721 \\[0.3mm]
$E\p$ & s & 5 & 2 & 1 & $E\pp$ & a & 4 & 2 & 1 & 1431137 & 0.2333$\times 10^{-1}$ & -0.717 \\[0.3mm]
$A_2\pp$ & s & 5 & 3 & 1 & $A_2\p$ & a & 4 & 3 & 1 & 1429410 & 0.1768$\times 10^{-1}$ & -0.710 \\[0.3mm]
$A_1\pp$ & s & 5 &-3 & 1 & $A_1\p$ & a & 4 &-3 & 1 & 1429409 & 0.1768$\times 10^{-1}$ & -0.710 \\[0.3mm]
$E\p$ & s & 5 & 4 & 1 & $E\pp$ & a & 4 & 4 & 1 & 1426908 & 0.9864$\times 10^{-2}$ & -0.700 \\[1mm]
 &  &   &   &   &   &   & &$^{15}$ND$_3$  &   &         &           &       \\[1mm]
$A_1\pp$ & a & 1 & 0 & 1 & $A_1\p$ & s & 0 & 0 & 1 &  402779 & 0.4636$\times 10^{-3}$ & -1.979 \\[0.3mm]
$A_2\pp$ & a & 2 & 0 & 1 & $A_2\p$ & s & 1 & 0 & 1 &  707552 & 0.2995$\times 10^{-2}$ & -1.551 \\[0.3mm]
$E\p$ & a & 2 & 1 & 1 & $E\pp$ & s & 1 & 1 & 1 &  708033 & 0.2250$\times 10^{-2}$ & -1.554 \\[0.3mm]
$A_2\p$ & s & 1 & 0 & 1 & $A_2\pp$ & a & 0 & 0 & 1 &  208813 & 0.6139$\times 10^{-4}$ &  0.891 \\[0.3mm]
$A_1\p$ & s & 2 & 0 & 1 & $A_1\pp$ & a & 1 & 0 & 1 &  515358 & 0.1131$\times 10^{-2}$ & -0.241 \\[0.3mm]
$E\pp$ & s & 2 & 1 & 1 & $E\p$ & a & 1 & 1 & 1 &  514961 & 0.8458$\times 10^{-3}$ & -0.237 \\[1mm]
\hline \\[-3mm]
%\hline \\[-1mm]
\end{tabular}
\vspace*{1mm}
\\
\footnotesize Frequencies from \citet{Bester}. \\
\end{table*}

\begin{table*}
\vspace*{-0.0cm}
\caption{The wavenumbers ($\nu$), wavelengths ($\lambda$), Einstein coefficients ($A$), and sensitivities ($T$) for transitions between the ground and $\nu_2$ vibrational state of $^{14}$ND$_3$ and $^{15}$ND$_3$.}
\label{tab:v2_14nd3_15nd3_2}
%\begin{ruledtabular}
\resizebox{\linewidth}{!}{\begin{tabular}{c @{\extracolsep{0.01in}}
                c @{\extracolsep{0.01in}}
                c @{\extracolsep{0.01in}}
                c @{\extracolsep{0.01in}}
                c @{\extracolsep{0.1in}}
                c @{\extracolsep{0.01in}}
                c @{\extracolsep{0.01in}}
                c @{\extracolsep{0.01in}}
                c @{\extracolsep{0.01in}}
                c @{\extracolsep{0.2000in}}
                c @{\extracolsep{0.1500in}}
                c @{\extracolsep{0.1500in}}
                c @{\extracolsep{0.1500in}}
                c @{\extracolsep{0.0500in}}
                c}
%& & & & & & & &  \\
\hline \\[-2mm]
\multicolumn{1}{c}{$\Gamma\p$}&\multicolumn{1}{c}{$p\p$}&\multicolumn{1}{c}{$J\p$}&\multicolumn{1}{c}{$K\p$}&\multicolumn{1}{c}{$v_{2}\p$}&\multicolumn{1}{c}{$\Gamma\pp$}&
\multicolumn{1}{c}{$p\pp$}&\multicolumn{1}{c}{$J\pp$}&\multicolumn{1}{c}{$K\pp$}&\multicolumn{1}{c}{$v_{2}\pp$}&
\multicolumn{1}{c}{$\nu$/cm$^{-1}$}&\multicolumn{1}{c}{$\lambda$/$\mu$m}&\multicolumn{1}{c}{$A$/s$^{-1}$}&\multicolumn{1}{c}{$T$}\\[1mm]
\hline \\[-1mm]
  &  &   &   &   &   &  & & & $^{14}$ND$_3$     &  &         &                 \\[1mm]
$A_1\pp$ & a &  1 &  0 & 1 & $A_1\p$ & s &  0 &  0 & 0 &   759.3704 & 13.1688 & 0.1955$\times 10^{+1}$ & -0.475  \\[0.3mm]
$A_2\pp$ & a &  2 &  0 & 1 & $A_2\p$ & s &  1 &  0 & 0 &   769.5283 & 12.9950 & 0.2444$\times 10^{+1}$ & -0.482  \\[0.3mm]
$E\p$ & a &  2 &  1 & 1 & $E\pp$ & s &  1 &  1 & 0 &   769.5306 & 12.9949 & 0.1834$\times 10^{+1}$ & -0.482  \\[0.3mm]
$A_2\p$ & s &  1 &  0 & 1 & $A_2\pp$ & a &  0 &  0 & 0 &   755.7906 & 13.2312 & 0.1948$\times 10^{+1}$ & -0.454  \\[0.3mm]
$A_1\p$ & s &  2 &  0 & 1 & $A_1\pp$ & a &  1 &  0 & 0 &   765.9901 & 13.0550 & 0.2434$\times 10^{+1}$ & -0.461  \\[0.3mm]
$E\pp$ & s &  2 &  1 & 1 & $E\p$ & a &  1 &  1 & 0 &   765.9767 & 13.0552 & 0.1827$\times 10^{+1}$ & -0.461  \\[0.3mm]
$E\p$ & a &  1 &  1 & 1 & $E\pp$ & s &  1 &  1 & 0 &   749.0866 & 13.3496 & 0.2810$\times 10^{+1}$ & -0.468  \\[0.3mm]
$E\p$ & a &  2 &  1 & 1 & $E\pp$ & s &  2 &  1 & 0 &   748.9645 & 13.3518 & 0.9344$\times 10^{0}$ & -0.468  \\[0.3mm]
$E\pp$ & a &  2 &  2 & 1 & $E\p$ & s &  2 &  2 & 0 &   748.9671 & 13.3517 & 0.3744$\times 10^{+1}$ & -0.468  \\[0.3mm]
$E\pp$ & s &  1 &  1 & 1 & $E\p$ & a &  1 &  1 & 0 &   745.4912 & 13.4140 & 0.2798$\times 10^{+1}$ & -0.446  \\[0.3mm]
$E\pp$ & s &  2 &  1 & 1 & $E\p$ & a &  2 &  1 & 0 &   745.4112 & 13.4154 & 0.9305$\times 10^{0}$ & -0.446  \\[0.3mm]
$E\p$ & s &  2 &  2 & 1 & $E\pp$ & a &  2 &  2 & 0 &   745.3664 & 13.4162 & 0.3729$\times 10^{+1}$ & -0.446  \\[0.3mm]
$A_2\pp$ & a &  0 &  0 & 1 & $A_2\p$ & s &  1 &  0 & 0 &   738.8622 & 13.5343 & 0.5381$\times 10^{+1}$ & -0.461  \\[0.3mm]
$A_1\pp$ & a &  1 &  0 & 1 & $A_1\p$ & s &  2 &  0 & 0 &   728.5209 & 13.7264 & 0.3427$\times 10^{+1}$ & -0.453  \\[0.3mm]
$E\p$ & a &  1 &  1 & 1 & $E\pp$ & s &  2 &  1 & 0 &   728.5205 & 13.7264 & 0.2572$\times 10^{+1}$ & -0.453  \\[0.3mm]
$A_1\p$ & s &  0 &  0 & 1 & $A_1\pp$ & a &  1 &  0 & 0 &   735.2618 & 13.6006 & 0.5358$\times 10^{+1}$ & -0.439  \\[0.3mm]
$A_2\p$ & s &  1 &  0 & 1 & $A_2\pp$ & a &  2 &  0 & 0 &   724.9421 & 13.7942 & 0.3412$\times 10^{+1}$ & -0.431  \\[0.3mm]
$E\pp$ & s &  1 &  1 & 1 & $E\p$ & a &  2 &  1 & 0 &   724.9258 & 13.7945 & 0.2560$\times 10^{+1}$ & -0.431 \\[1mm]
  &  &   &   &   &   &  & & & $^{15}$ND$_3$     &  &         &                 \\[1mm]
$A_1\pp$ & a &  1 &  0 & 1 & $A_1\p$ & s &  0 &  0 & 0 &   752.9702 & 13.2807 & 0.1888$\times 10^{+1}$ & -0.475  \\[0.3mm]
$A_2\pp$ & a &  2 &  0 & 1 & $A_2\p$ & s &  1 &  0 & 0 &   763.1000 & 13.1044 & 0.2359$\times 10^{+1}$ & -0.482  \\[0.3mm]
$E\p$ & a &  2 &  1 & 1 & $E\pp$ & s &  1 &  1 & 0 &   763.0998 & 13.1044 & 0.1770$\times 10^{+1}$ & -0.482  \\[0.3mm]
$A_2\p$ & s &  1 &  0 & 1 & $A_2\pp$ & a &  0 &  0 & 0 &   749.6973 & 13.3387 & 0.1881$\times 10^{+1}$ & -0.455  \\[0.3mm]
$A_1\p$ & s &  2 &  0 & 1 & $A_1\pp$ & a &  1 &  0 & 0 &   759.8667 & 13.1602 & 0.2351$\times 10^{+1}$ & -0.463  \\[0.3mm]
$E\pp$ & s &  2 &  1 & 1 & $E\p$ & a &  1 &  1 & 0 &   759.8517 & 13.1605 & 0.1764$\times 10^{+1}$ & -0.462  \\[0.3mm]
$E\p$ & a &  1 &  1 & 1 & $E\pp$ & s &  1 &  1 & 0 &   742.7222 & 13.4640 & 0.2713$\times 10^{+1}$ & -0.468  \\[0.3mm]
$E\p$ & a &  2 &  1 & 1 & $E\pp$ & s &  2 &  1 & 0 &   742.6101 & 13.4660 & 0.9023$\times 10^{0}$ & -0.468  \\[0.3mm]
$E\pp$ & a &  2 &  2 & 1 & $E\p$ & s &  2 &  2 & 0 &   742.6053 & 13.4661 & 0.3616$\times 10^{+1}$ & -0.468  \\[0.3mm]
$E\pp$ & s &  1 &  1 & 1 & $E\p$ & a &  1 &  1 & 0 &   739.4346 & 13.5238 & 0.2702$\times 10^{+1}$ & -0.448  \\[0.3mm]
$E\pp$ & s &  2 &  1 & 1 & $E\p$ & a &  2 &  1 & 0 &   739.3626 & 13.5252 & 0.8988$\times 10^{0}$ & -0.448  \\[0.3mm]
$E\p$ & s &  2 &  2 & 1 & $E\pp$ & a &  2 &  2 & 0 &   739.3131 & 13.5261 & 0.3602$\times 10^{+1}$ & -0.447  \\[0.3mm]
$A_2\pp$ & a &  0 &  0 & 1 & $A_2\p$ & s &  1 &  0 & 0 &   732.5333 & 13.6513 & 0.5197$\times 10^{+1}$ & -0.461  \\[0.3mm]
$A_1\pp$ & a &  1 &  0 & 1 & $A_1\p$ & s &  2 &  0 & 0 &   722.2354 & 13.8459 & 0.3311$\times 10^{+1}$ & -0.452  \\[0.3mm]
$E\p$ & a &  1 &  1 & 1 & $E\pp$ & s &  2 &  1 & 0 &   722.2324 & 13.8460 & 0.2484$\times 10^{+1}$ & -0.453  \\[0.3mm]
$A_1\p$ & s &  0 &  0 & 1 & $A_1\pp$ & a &  1 &  0 & 0 &   729.2409 & 13.7129 & 0.5176$\times 10^{+1}$ & -0.440  \\[0.3mm]
$A_2\p$ & s &  1 &  0 & 1 & $A_2\pp$ & a &  2 &  0 & 0 &   718.9634 & 13.9089 & 0.3296$\times 10^{+1}$ & -0.432  \\[0.3mm]
$E\pp$ & s &  1 &  1 & 1 & $E\p$ & a &  2 &  1 & 0 &   718.9456 & 13.9093 & 0.2474$\times 10^{+1}$ & -0.432  \\[1mm]
\hline \\[-3mm]
%\\
\end{tabular}}
\vspace*{1mm}
\\
\footnotesize Wavenumbers from \citet{Bester}. \\
%\end{ruledtabular}
\end{table*}

It is expected that any variation in the fundamental constants will be confirmed, or refuted, over a series of independent measurements on a variety of molecular absorbers. As a relevant astrophysical molecule, and with certain inversion transitions already detected extraterrestrially~\citep{Mauersberger:1986,Johnston:1989,Schilke:1991}, $^{15}$NH$_3$ has potential to aid this search along with the already established probes of $^{14}$NH$_3$. For the deuterated species $^{14}$ND$_3$ and $^{15}$ND$_3$, it is perhaps more likely that their use will be restricted to precision measurements in the laboratory, despite possessing larger sensitivity coefficients for the pure inversion frequencies in the ground vibrational state.

\section{Conclusion}

A comprehensive study of the vibration-rotation-inversion transitions of all stable, symmetric top isotopomers of ammonia has been performed. The variational method offers a new and robust approach to computing sensitivity coefficients. The calculated mass sensitivities provide perspectives for the further development of the ammonia method, used in the probing of the cosmological variability of the proton-to-electron mass ratio. Most notably the reliance on other reference molecular species, which is the main source of systematic error, can be avoided. Although ammonia is not a primordial molecule and cannot be studied at extremal redshifts such as H$_2^+$, D$_2^+$ and He$_2^+$ for instance~\citep{Lucie:2014}, it can be detected in a wide variety of regions~\citep{Ho:1983}, and at redshifts which dramatically enhance spectral shifts (see \citet{Riechers:2013} and also Eq.~(1) of \citet{Spirko:2014}). The accuracy of the predicted sensitivities seems to fulfil the requirements needed for a reliable analysis of spectral data obtained at `rotational' resolution. To go beyond this limit, one should account for the hyperfine interactions and this requires a correct description of the `hyperfine' effects, which in turn should respect both the centrifugal distortion and Coriolis interaction~\citep{Patrick:1990}. A study along these lines is in progress in our laboratory. We also note that the ammonia rovibrational dynamics show the same characteristics as those of other inverting molecules, notably the hydronium cation (see \citet{Kozlov:2011}), thus calling for a rigorous investigation into such systems.

\section*{Acknowledgements}
The work was a part of the research project RVO:61388963 (IOCB) and was supported by the Czech Science Foundation (grant P209/15-10267S). S.Y. thanks ERC Advanced Investigator Project 267219 and A.O. thanks the UCL Impact scheme. The authors are grateful to Luciano Fusina, Gianfranco Di Lonardo and Adriana Predoi-Cross for providing their $^{15}$NH$_3$ data prior to publishing.

\singlespacing
\bibliographystyle{apsrev}
\bibliography{ammonia_probes}

\label{lastpage}

\end{document}